\documentclass[fleqn,usenatbib]{mnras}

\usepackage{newtxtext,newtxmath}

\usepackage[T1]{fontenc}

\DeclareRobustCommand{\VAN}[3]{#2}
\let\VANthebibliography\thebibliography
\def\thebibliography{\DeclareRobustCommand{\VAN}[3]{##3}\VANthebibliography}

\usepackage{graphicx}	
\usepackage{amsmath}	

\usepackage{caption}
\usepackage{subcaption}
\usepackage{booktabs}
\usepackage[section]{placeins}
\usepackage{float}
\usepackage{tablefootnote}

\title[Resonances 1/-2, 2/-1, 1/-1 in planetary systems]{A numerical study of the 1/2, 2/1 and 1/1 retrograde mean motion resonances in planetary systems}

\author[G. A. Caritá et al.]{
Gabriel António Caritá,$^{1}$\thanks{E-mail: gabrielcarita@gmail.com}
Alan Cefali Signor,$^{1}$
Maria Helena Moreira Morais,$^{1}$\thanks{E-mail: helena.morais@unesp.br}
\\
$^{1}$Instituto de Geociências e Ciências Exatas, Universidade Estadual Paulista (UNESP), Av. 24-A, 1515, 13506-900 Rio Claro, SP, Brazil \\
}

\date{Accepted 08 Jun 2022. Received 11 May 2022; in original form ZZZ}

\pubyear{2022}

\begin{document}
\label{firstpage}
\pagerange{\pageref{firstpage}--\pageref{lastpage}}
\maketitle

\begin{abstract}
We present a numerical study on the stability of the 1/2, 2/1 and 1/1 retrograde mean motion resonances in the 3-body problem composed of a solar mass star, a  Jupiter mass planet and an additional body with zero mass (elliptic restricted 3-body problem) or  masses corresponding to either Neptune, Saturn or Jupiter (planetary 3-body problem). For each system we obtain stability maps  using the n-body numerical integrator REBOUND and  computing the chaos indicator \textit{mean exponential growth factor of nearby orbits} (MEGNO). We show that families of periodic orbits exist in all configurations and they correspond to the libration of either a single resonant argument  or all resonant arguments (fixed points). We compare the results obtained in the elliptic restricted 3-body problem with previous results in the literature and we show the differences and similarities between the phase space topology for these retrograde resonances in the circular restricted, elliptic restricted and planetary 3-body problems.  
\end{abstract}




\section{Introduction}

A retrograde resonance configuration occurs when 2 objects orbit a star in opposite directions  and their orbital frequencies are commensurable \citep{morais2012stability}.  
The study of retrograde resonance in the framework of the circular restricted 3-body problem \citep{2012DPS....4411222D,morais2013retrograde,namouni2015resonance,morais2016numerical}, allowed the identification of the first small bodies in retrograde resonances in the solar system, namely Centaurs in retrograde resonances with Jupiter and Saturn \citep{morais2013asteroids}  and asteroid (514107) Ka`epaoka`awela in the 1/1 (co-orbital) retrograde resonance with Jupiter \citep{wiegert2017retrograde,morais2017reckless,namouni2018coorbital}. More recently, the families of periodic orbits which are associated with retrograde resonances have been computed in the spacial restricted 3-body problem \citep{morais2019periodic,kotoulas2020retrograde,moraisetal2021,kotoulas2022three} and elliptic restricted 3-body problem \citep{kotoulas2020planar,kotoulas2020retrograde,kotoulas2022three}

The possibility of retrograde resonances in extra-solar systems has been proposed by \cite{gayon2008retrograde,gayon2009fitting} but no exhaustive studies on the stability of such configurations had been conducted until now. The formation of systems with counter-revolving planets is possible e.g.\  due to  close encounters between  stars leading to exchange of planets between them \citep{malmberg2011effects}. These captured planets may have high inclination orbits with respect to the ones that formed in situ.

In this article we present a numerical investigation of retrograde resonances in the planar planetary three body problem using stability  and resonance maps. In Sect.~2 we explain our numerical search methodology and we present results for the retrograde resonances 1/2, 2/1 and 1/1 in the elliptic and planetary three body problems. In Sect.~3 we discuss these results and in Sect.~4 we present our conclusions.   

\section{Numerical study of retrograde mean motion resonances}

We consider  the  three body  problem composed of a solar mass star  and two planets  orbiting the star in opposite directions, one counter-clockwise (prograde) and the other clockwise (retrograde). We  use the notation $p/-q$ mean motion resonance to refer to a configuration where 2 planets  orbiting the star in opposite directions have mean motions (average orbital frequencies) which are nearly commensurable in the the ratio $p/q$ \citep{morais2013retrograde}. We use the usual notation for astrocentric orbital elements: $a$ (semi-major axis), $T$ (orbital period), $e$ (eccentricity), $I$ (inclination), $M$ (mean anomaly), $\omega$ (argument of pericenter), $\Omega$ (longitude of ascending node), $\lambda$ (mean longitude), $\varpi$ (longitude of pericenter). Variables with subscript $p$ refer to the prograde planet  and those without subscript refer to the retrograde planet. The prograde planet has mass $0.001\,M_\odot$ and the unit distance is its mean distance to the star, i.e.\ $a_p=1$. In the elliptic restricted 3-body problem (ER3BP) the retrograde planet has zero mass, and in the planetary 3-body problem it has mass equal to either Neptune ($0.00005149\,M_\odot$), Saturn ($0.0002857\,M_\odot$) or Jupiter ($0.001\,M_\odot$). We  investigate the stable configurations of planar systems in the  1/-2, 2/-1 and 1/-1 mean motion resonances.

The numerical  integration of the equations of motion were performed using REBOUND with the adaptive step integrator IAS15 \citep{rein2015ias15} and in some cases (especially high eccentricity orbits) using the Bulirsch-Stoer method implemented in MERCURY \citep{chambers1999hybrid}. The integration for a particular object was stopped when the mutual distance to another object was smaller that the sum of their radii (collision) or when the distance to the star was larger than $10\,a_p$ (escape). 

We assume a counter-clockwise reference frame. The longitudes defined in the orbital plane for the prograde planet (with counter-clockwise motion) are measured in the direction of the object's orbital motion, hence $\lambda_p = \Omega_p + \omega_p +M_p$, $\varpi_p = \Omega_p + \omega_p$, while those for the retrograde planet  (with clockwise motion) are measured against the  direction of the object's orbital motion, hence $\lambda = \Omega - \omega -M$, $\varpi = \Omega - \omega$. This is also the the convention used within REBOUND.

We construct stability  and resonant maps. The stability maps are obtained by computation of the chaotic indicator MEGNO, which converges to 2 for stable orbits while values greater than 2 indicate chaos \citep{cincotta2000simple,gozdziewski2003stability}. The resonant maps indicate regions of libration of the resonant angle of the circular restricted 3-body problem which  for a $p/-q$ mean motion resonance is  $\phi_{0}= -q\lambda - p\lambda_p + (p+q)\varpi$ \citep{morais2013retrograde}. There are also regions  of libration of both $\phi_0$ and $\varpi-\varpi_p$ (which indicates a fixed point of the resonant problem), identified by a white symbol. Regions of libration of a single resonant argument different from $\phi_0$ (which we will define shortly for each resonance) are identified by a black symbol.
The color bar in the resonant maps indicates the semi-amplitude of $\phi_0$ libration i.e.\ the maximum variation around the resonant center. We identify  fixed points when both $\phi_0$ and $\varpi-\varpi_p$ librate (around either $0$ or $\pi$) with semi-amplitude less than $\pi/4$.

We construct 2 types of maps: 1) eccentricity of the prograde planet versus eccentricity of the retrograde planet with initial semi-major axis fixed at the nominal resonance location $a=(q/p)^{2/3}$; 2)  eccentricity versus semi-major axis of the retrograde planet at fixed initial prograde planet's eccentricity. In 1) we used a grid of 80$\times$80  initial eccentricities in the range $[0,1]$ while in 2) we have initial $(a,e)$ in a 40$\times$40 (low resolution) or 80$\times$80 grid (high resolution). In both cases we integrated for $2\,\times 10^5\,T_p$.

We set the initial longitudes of the nodes to be zero ($\Omega=\Omega_p=0$). The initial inclination of the retrograde planet with respect to the prograde planet's orbital plane is set to $I=179.99^\circ$. In some cases we will see differences between this nearly-coplanar case with results presented for strictly 2D cases (where the zed components of the positions and velocities of all objects are strictly zero). In practice this means that such 2D periodic orbits are vertically unstable and will not exist in real systems. Therefore, integrating the nearly co-planar case is more relevant for studying the dynamics of real systems which will never be exactly coplanar.

\begin{figure*}
\centering
\begin{subfigure}[b]{0.85\textwidth}
\centering
\includegraphics[width=\textwidth]{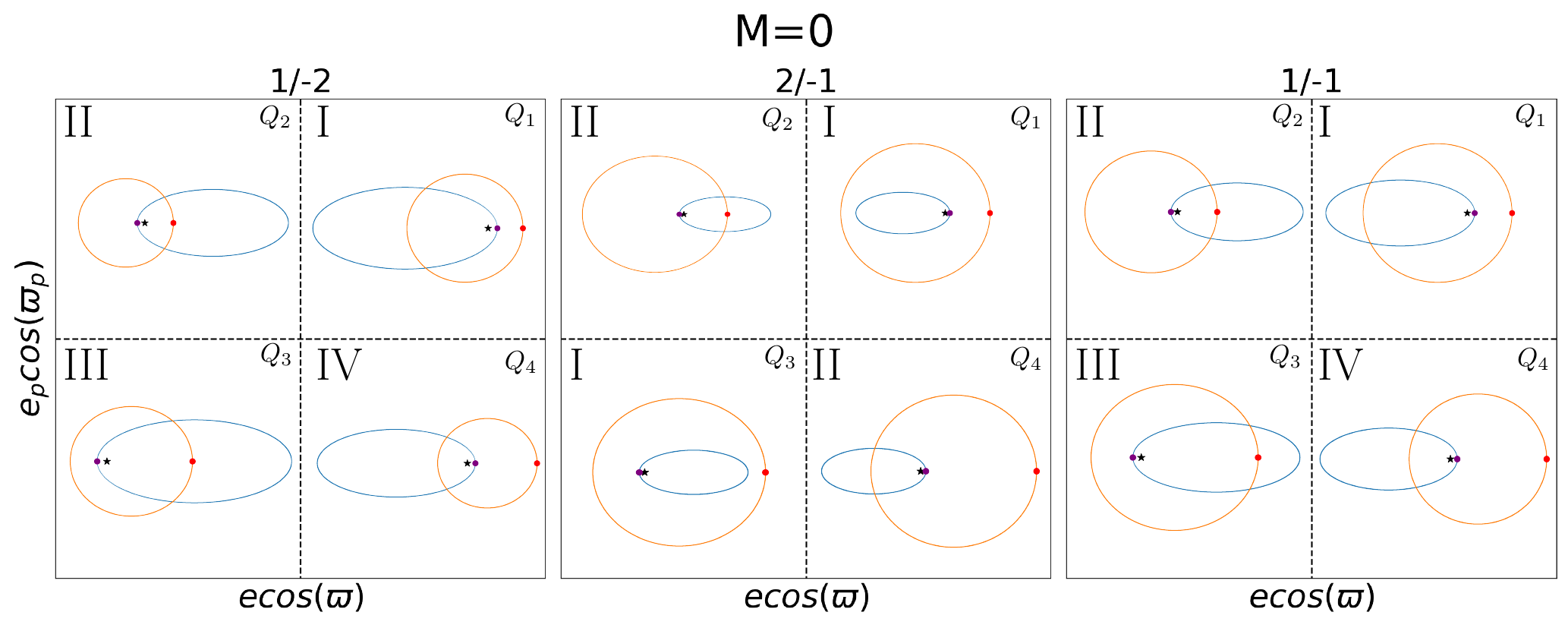}
\caption{$M=0$}
\label{fig:configurationsm0}
\end{subfigure}
\vskip15pt
\begin{subfigure}[b]{0.85\textwidth}
\centering
\includegraphics[width=\textwidth]{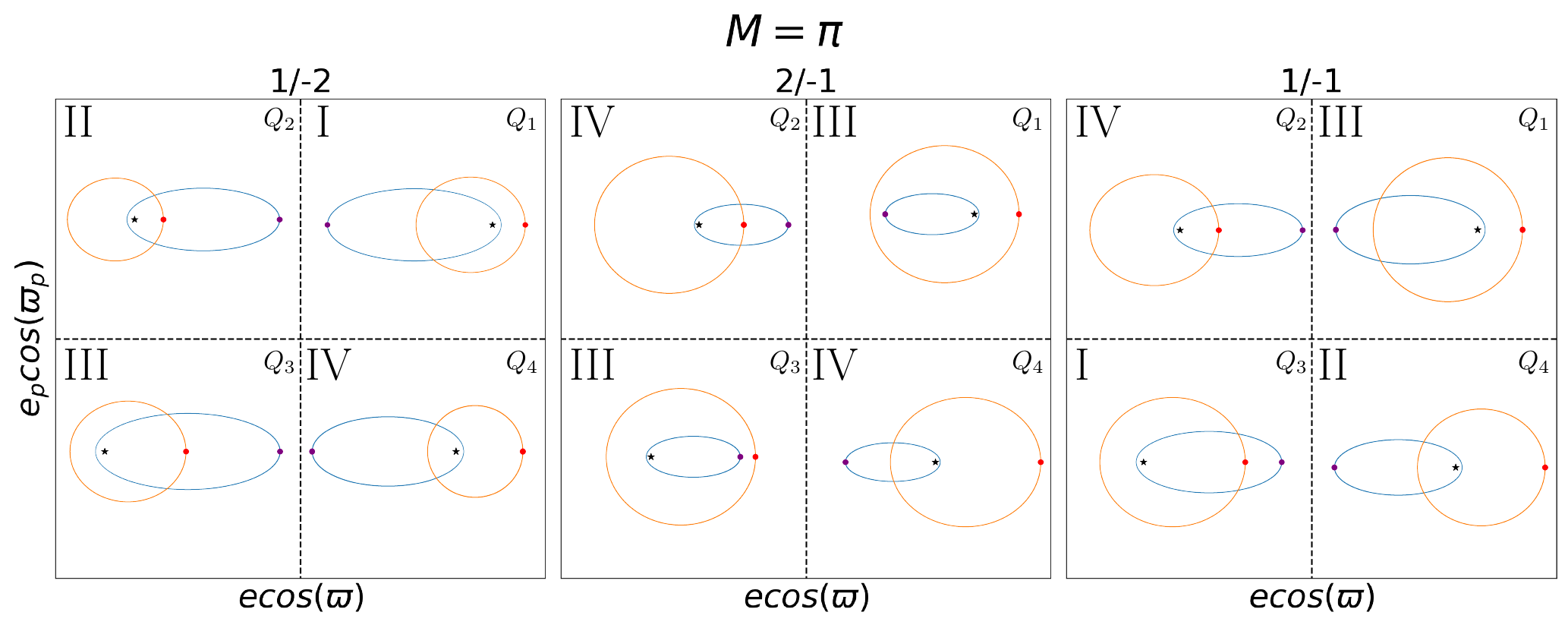}
\caption{$M=\pi$}
\label{fig:configurationsm180}
\end{subfigure}
\caption{Configurations displayed in 4 quadrants which represent the initial orientations of the pericenters / apocenters, and initial 
conditions $M_p=\varpi_p$ and $M=0$ (a) or $M=\pi$ (b). The Roman numerals indicate pairing of initial configurations which, due to the commensurability between the orbital periods, are equivalent with a time-lag of half a period of the external object.}
\label{fig:configurations}
\end{figure*}

We investigate all possible permutations of aligned / anti-aligned pericenters / apocenters  separated by quadrants displayed in Figure \ref{fig:configurations}. In $Q_1$,  $\varpi_p=\varpi=0$; in $Q_2$,  $\varpi_p=0$, $\varpi=\pi$; in $Q_3$, $\varpi_p=\varpi=\pi$; in $Q_4$, $\varpi_p=\pi$,$\varpi=0$. The initial mean anomaly of the prograde planet is $M_p = \varpi_p$, while the mean anomaly of the retrograde body is  $M=0$ (Figure \ref{fig:configurations} (a)), or $M=\pi$ (Figure \ref{fig:configurations} (b)). The Roman numerals indicate pairing of initial configurations which, due to the commensurability between the orbital periods, are equivalent with a time-lag of half a period of the external object. This equivalence is not exact due to the interaction between the planets during that time-lag.

\subsection{1/-2 Resonance}

For resonance 1/-2 the resonant angles analyzed were:

\begin{equation}
    \phi_{0} = -2\lambda - \lambda_p + 3\varpi
\end{equation}

\begin{equation}
    \phi_1 = -2\lambda - \lambda_p + 3\varpi_p
\end{equation}

\begin{equation}
   \phi_2 = -2\lambda - \lambda_p + 2\varpi_p + \varpi
\end{equation}

\begin{equation}
    \phi_3 = -2\lambda - \lambda_p + \varpi_p + 2\varpi
\end{equation}

Table \ref{tabela1-2sum} presents the summarized results of the 1/-2 resonant configurations, indicating the object's masses and libration angles.   
\begin{table*}
\centering
\caption{Table reporting the summarized results for the 1/-2 resonance.
The notation $\phi_{0,1,2,3}$ indicates fixed point libration ($\phi_0$ and $\varpi-\varpi_p$ are fixed), and either $\phi_{0}$ or $\phi_3$ indicates libration of a single angle.}
\label{tabela1-2sum}
\resizebox{0.7\textwidth}{!}{%
\begin{tabular}{cccccc}
\hline
\textbf{Mass} & \multicolumn{4}{c}{Resonance} & \textbf{Figure} \\ \hline
              & $Q_1$  & $Q_2$ & $Q_3$ & $Q_4$ &                 \\ \hline
ER3BP($M=0$) &  $\phi_{0,1,2,3}$ and $\phi_{0}$ & $\phi_{0}$ & $\phi_{0}$ & $\phi_{0}$ & \ref{fig12:1er3bp}            \\
ER3BP($M=\pi$)   &  $\phi_{0,1,2,3}$ and $\phi_{0}$ & $\phi_{0}$ & $\phi_{0}$ & $\phi_{0}$ & \ref{fig12:2er3bp}              \\
NEPTUNE($M=0$)  & $\phi_{0,1,2,3}$ and $\phi_{0}$ & $\phi_{0}$ & $\phi_{0}$ & $\phi_{0}$ & \ref{fig12:1nep}            \\
NEPTUNE($M=\pi$) &     $\phi_{0,1,2,3}$ and $\phi_{0}$ & $\phi_{0}$ & $\phi_{0}$ & $\phi_{0}$ & \ref{fig12:2nep}            \\
SATURN($M=0$)   &      $\phi_{0,1,2,3}$ and $\phi_{3}$  & $\phi_{0}$ and $\phi_{3}$ & $\phi_{0,1,2,3}$  & $\phi_{0}$  &      \ref{fig12:1sat}           \\
SATURN($M=\pi$)  &      $\phi_{0,1,2,3}$ and $\phi_{3}$  & $\phi_{0}$ and $\phi_{3}$ & $\phi_{0,1,2,3}$  & $\phi_{0}$ &  \ref{fig12:2sat}               \\
JUPITER($M=0$)  &      $\phi_{0,1,2,3}$ and $\phi_{3}$ &  $\phi_{3}$ & $\phi_{0,1,2,3}$ & -  &   \ref{fig12:1jup}              \\
JUPITER($M=\pi$) &      $\phi_{0,1,2,3}$ and $\phi_{3}$ & $\phi_{0}$ and $\phi_{3}$ & $\phi_{0,1,2,3}$ and $\phi_{0}$ &            - & \ref{fig12:2jup}           \\ \bottomrule
\end{tabular}%

}

\end{table*}

\begin{figure}
     \centering
     \begin{subfigure}[b]{0.43\textwidth}
         \centering
         \includegraphics[width=\textwidth]{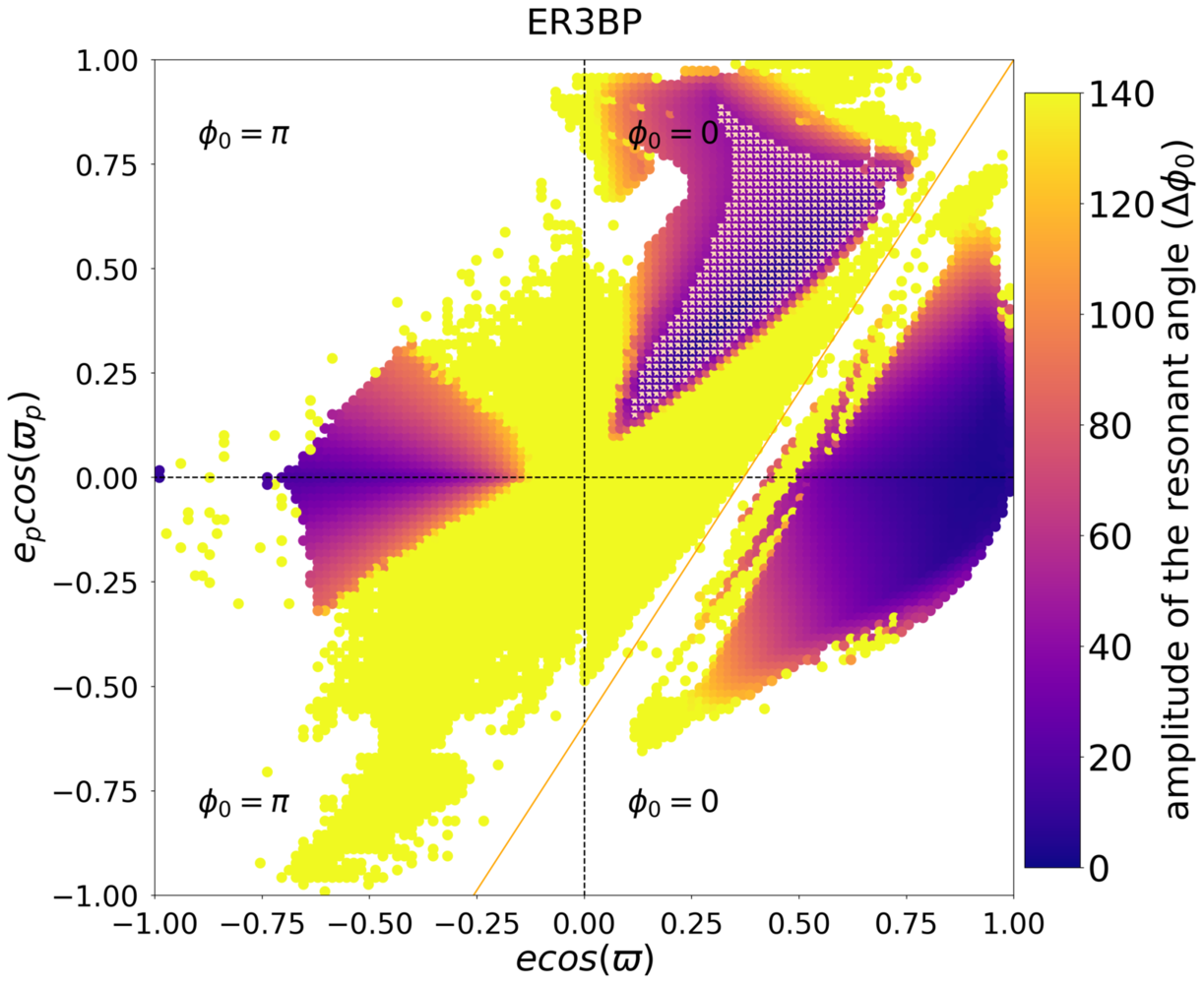}
         \caption{}
         \label{fig12:1er3bp}
     \end{subfigure}
     \vskip15pt
     \begin{subfigure}[b]{0.43\textwidth}
         \centering
         \includegraphics[width=\textwidth]{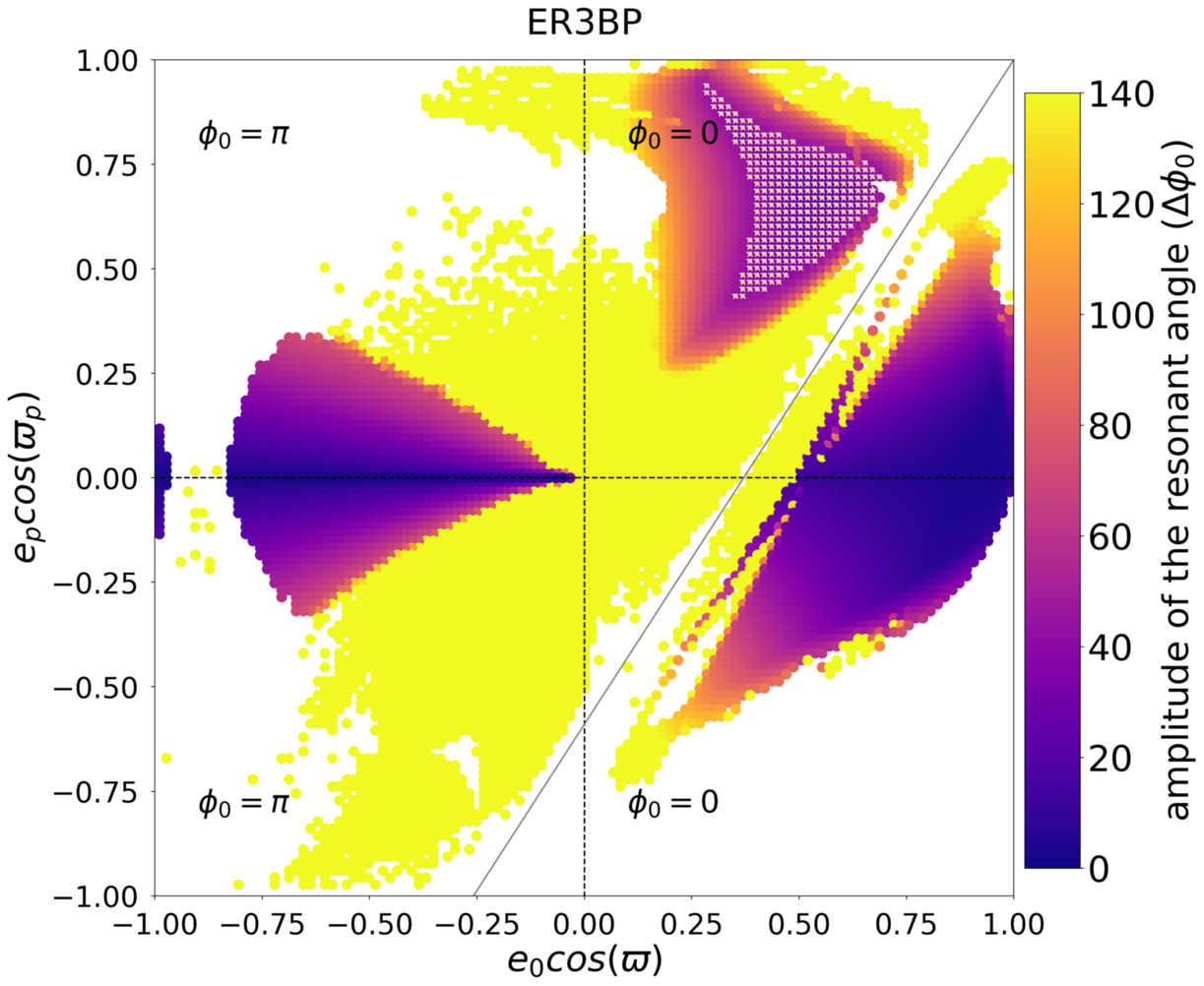}
         \caption{}
         \label{fig12:2er3bp}
     \end{subfigure}
     
     \caption{Resonant maps for the 1/-2 resonance in the elliptic restricted three body problem (a) $M=0$; (b) $M=\pi$. The color bar represents the amplitude of the restricted angle ($\phi_{0}$) and the overlaying white symbols indicate the fixed point family where all resonant angles librate around a center. The orange and gray lines indicate collision at time zero or after half a period of the external object, respectively.}
     \label{fig:1}
     \end{figure}

In Figure \ref{fig:1} for $M = 0$ (a) and   $M=\pi$ (b) we present the results for the ER3BP. The color bar indicates the amplitude of the resonant angle $\phi_{0}$, where dark purple/blue indicates the resonance center. In the 1st quadrant ($Q_1$) of the both maps there are periodic families where $\phi_0$ and $\varpi-\varpi_p$ (hence also  $\phi_1$, $\phi_2$, $\phi_3$)  librate simultaneously (fixed point families) which we represent by overlaying white symbols in the darker region. From Figure \ref{fig:configurations} we expect approximate symmetry between the quadrants in Figure \ref{fig:1} (a) and (b). In fact, we see the same structures in both panels but  for $M=\pi$ the fixed point family in $Q_1$ has decreased while the $\phi_{0}$ family has increased in quadrants $Q_2,Q_3$ and $Q_4$, in comparison with $M=0$.

The 1/-2 resonance at Jupiter to Sun mass ratio in the CR3BP was studied in \cite{morais2016retrograde}. In that work the $\phi_0 = 0$ family starts at $e > 0.4$ in agreement with our results (purple regions at $e_p=0$ on right hand side of \ref{fig:1} (a) and (b)), while the  $\phi_0 = \pi$ exists from $e=0$ up to $e\approx 0.8$  also in agreement with our results (purple regions at $e_p=0$ on left hand side of \ref{fig:1} (a) and (b)).
The 1/-2 resonance was also studied in the ER3BP at Neptune to Sun mass ratio in \cite{kotoulas2020retrograde}. The fixed point family which appears in the ER3BP at Jupiter to Sun mass ratio family (1st quadrants of \ref{fig:1} (a) and (b))  is qualitatively in agreement with the stable family computed by \cite{kotoulas2020retrograde}.

\begin{figure}
     \centering
     \begin{subfigure}[b]{0.43\textwidth}
         \centering
         \includegraphics[width=\textwidth]{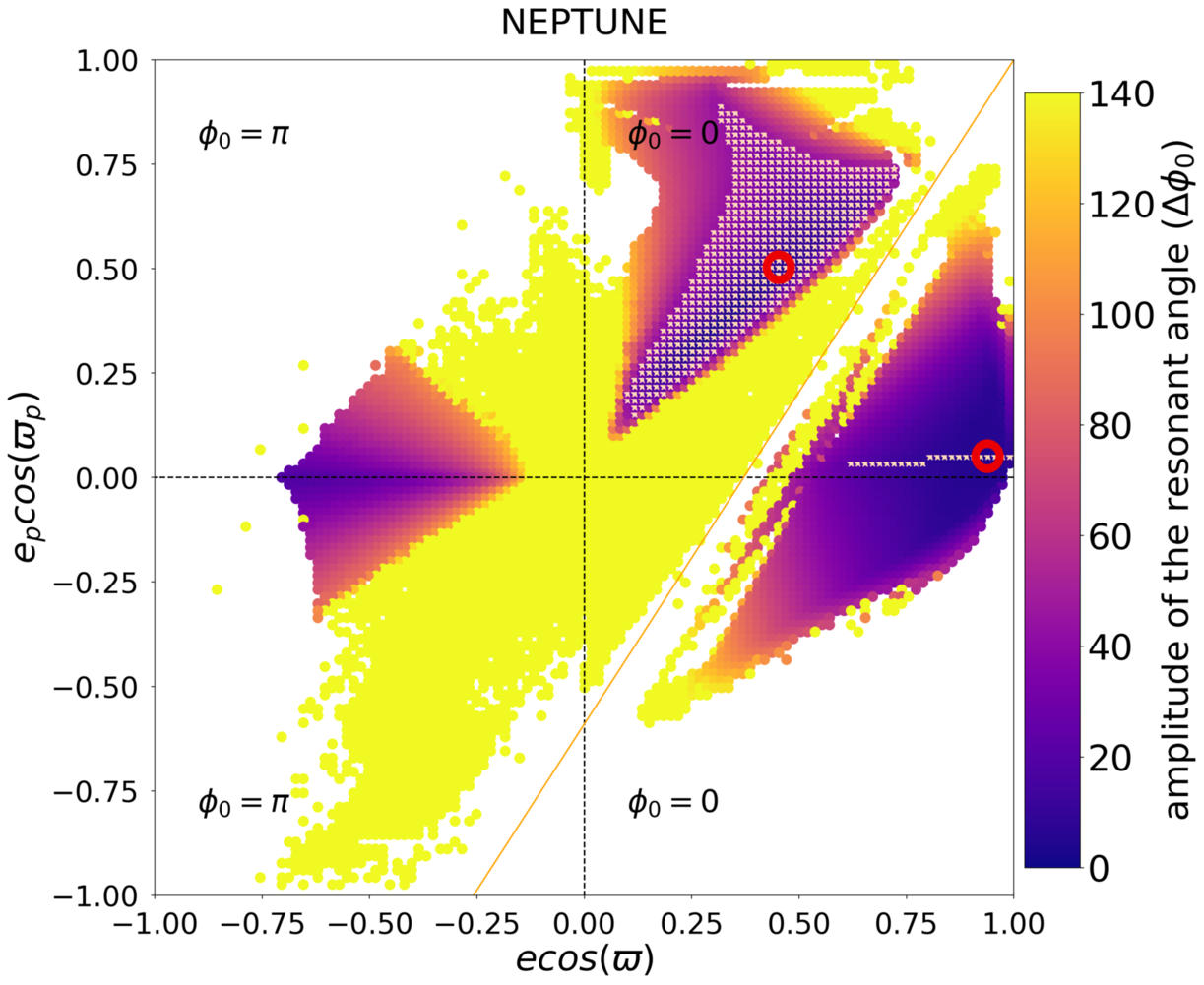}
         \caption{}
         \label{fig12:1nep}
     \end{subfigure}
     \vskip15pt
     \begin{subfigure}[b]{0.43\textwidth}
         \centering
         \includegraphics[width=\textwidth]{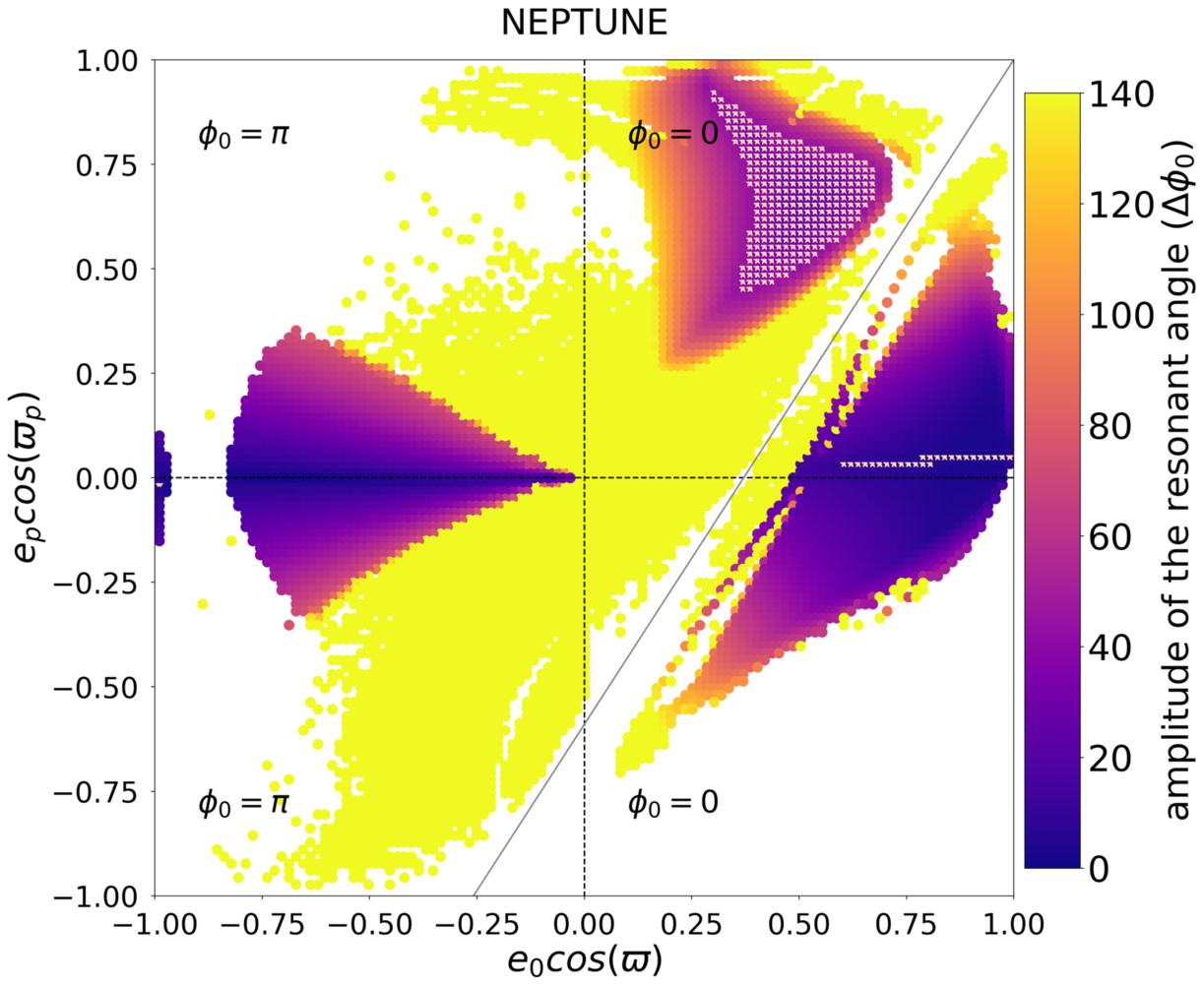}
         \caption{}
         \label{fig12:2nep}
     \end{subfigure}
     
     \caption{Resonant maps for the 1/-2 resonance in the planetary problem when the 2nd planet has Neptune's mass (a) $M=0$; (b) $M=\pi$. The color bar represents the amplitude of the restricted angle ($\phi_{0}$) and the overlaying white symbols indicate the fixed point family where all resonant angles librate around a center. The orange and gray lines indicate collision at time zero or after half a period of the external object, respectively.}
     \label{fig:2}
     \end{figure}

\begin{figure}
\centering
    \begin{subfigure}[b]{0.43\textwidth}
    \includegraphics[width=\textwidth]{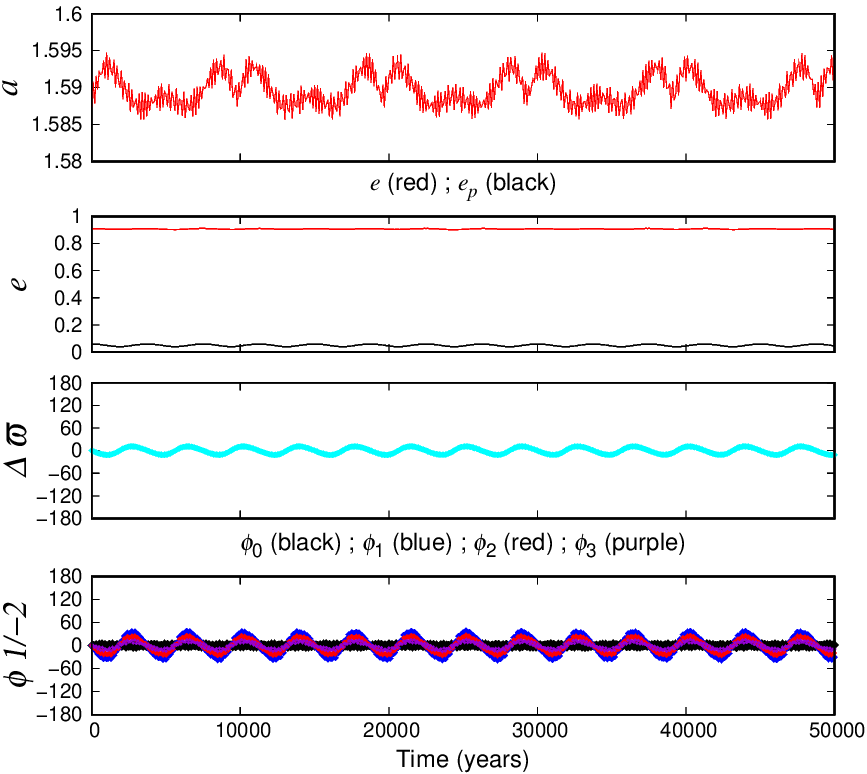}
    \caption{}
    \label{fig:12nepcia}
    \end{subfigure}
    \vskip15pt
    \begin{subfigure}[b]{0.43\textwidth}
      \includegraphics[width=\textwidth]{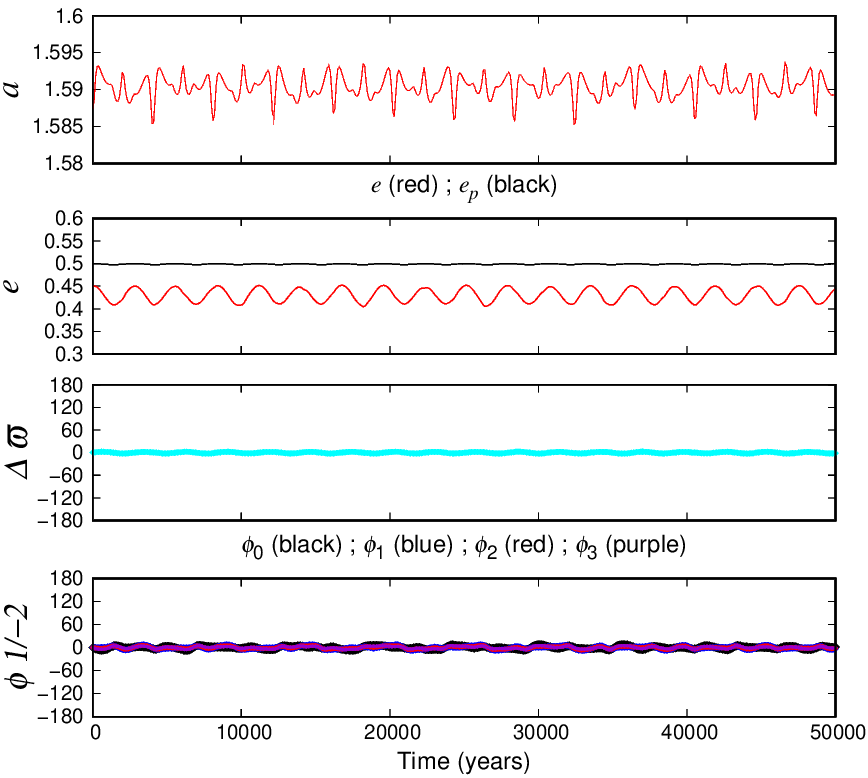}
    \caption{}
    \label{fig:12nepcib}
    \end{subfigure}
    \caption{Orbital evolution as a function of time for the initial conditions circled in Figure \ref{fig12:1nep} ($Q_1$, $M=0$).  In (a) the initial conditions are  $e = 0.91$ and $e_p = 0.058$. In (b) the initial conditions are  
    $e = 0.45$, $e_p = 0.5$.  In (a) and (b ) the 1st panel shows the semi-major axis of the third body, the 2nd panel shows  its orbital eccentricity, the 3rd panel shows  the difference $\Delta\varpi$ between the longitudes of pericenter, the 4th panel shows the resonant angles $\phi_0$, $\phi_1$, $\phi_2$, and $\phi_3$.}
\label{fig:12nepci}
\end{figure}

The stability maps for the planetary problem when the second planet has Neptune's mass, are presented Figure \ref{fig:2}. We can observe the same resonant families present in the ER3BP, however, a new fixed point resonance region  appears in the first quadrant, near $e_p = 0.05$. This happens for both cases $M=0$ and $M=\pi$. 
In Figure \ref{fig:12nepci} we show the orbital evolution corresponding to the initial conditions marked with red circles in Figure \ref{fig:2}. In both cases all resonant angles and $\varpi-\varpi_p$ librate around 0 with small amplitude as expected for a fixed point resonant family.  

\begin{figure}
     \centering
     \begin{subfigure}[b]{0.43\textwidth}
         \centering
         \includegraphics[width=\textwidth]{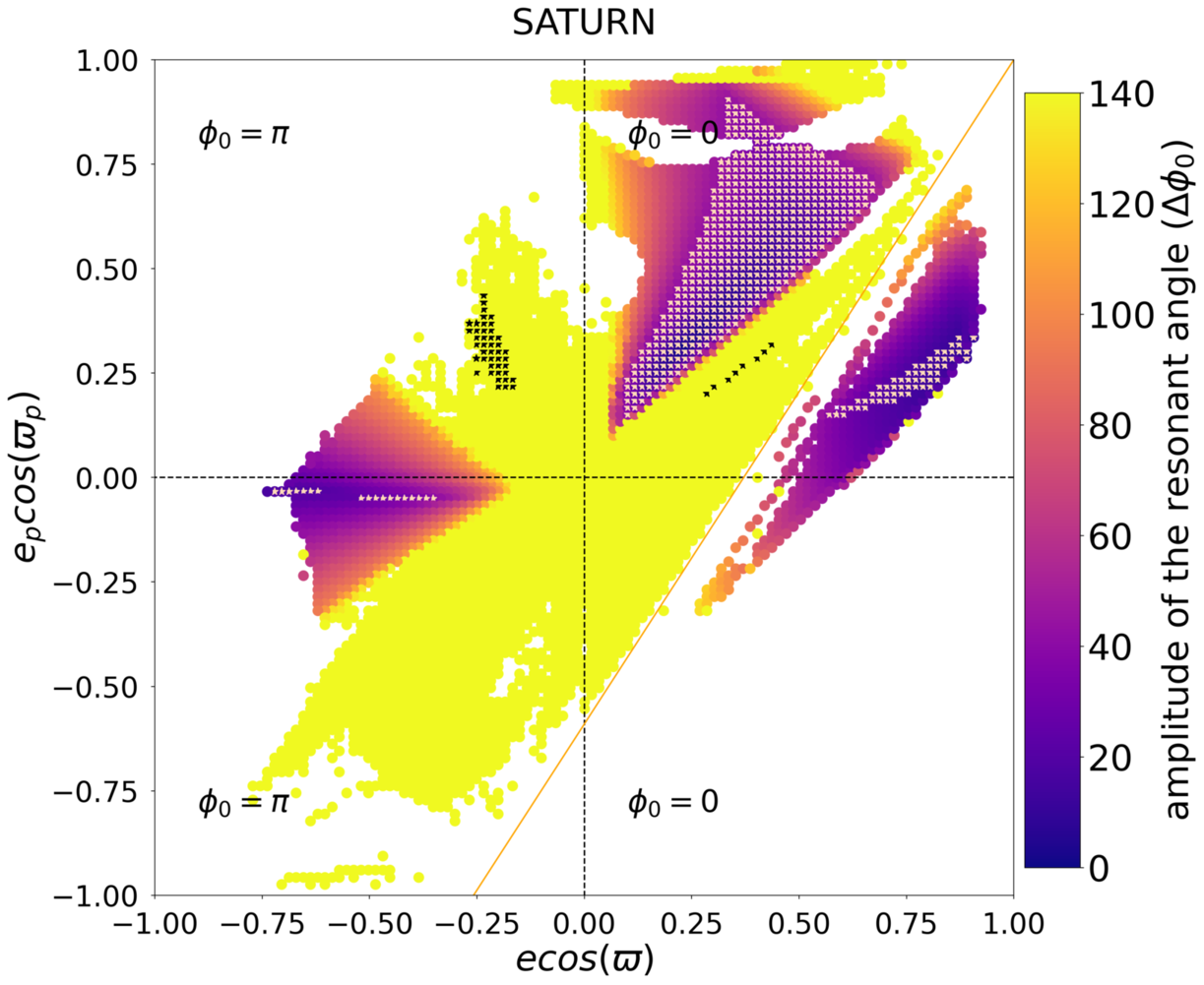}
         \caption{}
         \label{fig12:1sat}
     \end{subfigure}
     \vskip15pt
     \begin{subfigure}[b]{0.43\textwidth}
         \centering
         \includegraphics[width=\textwidth]{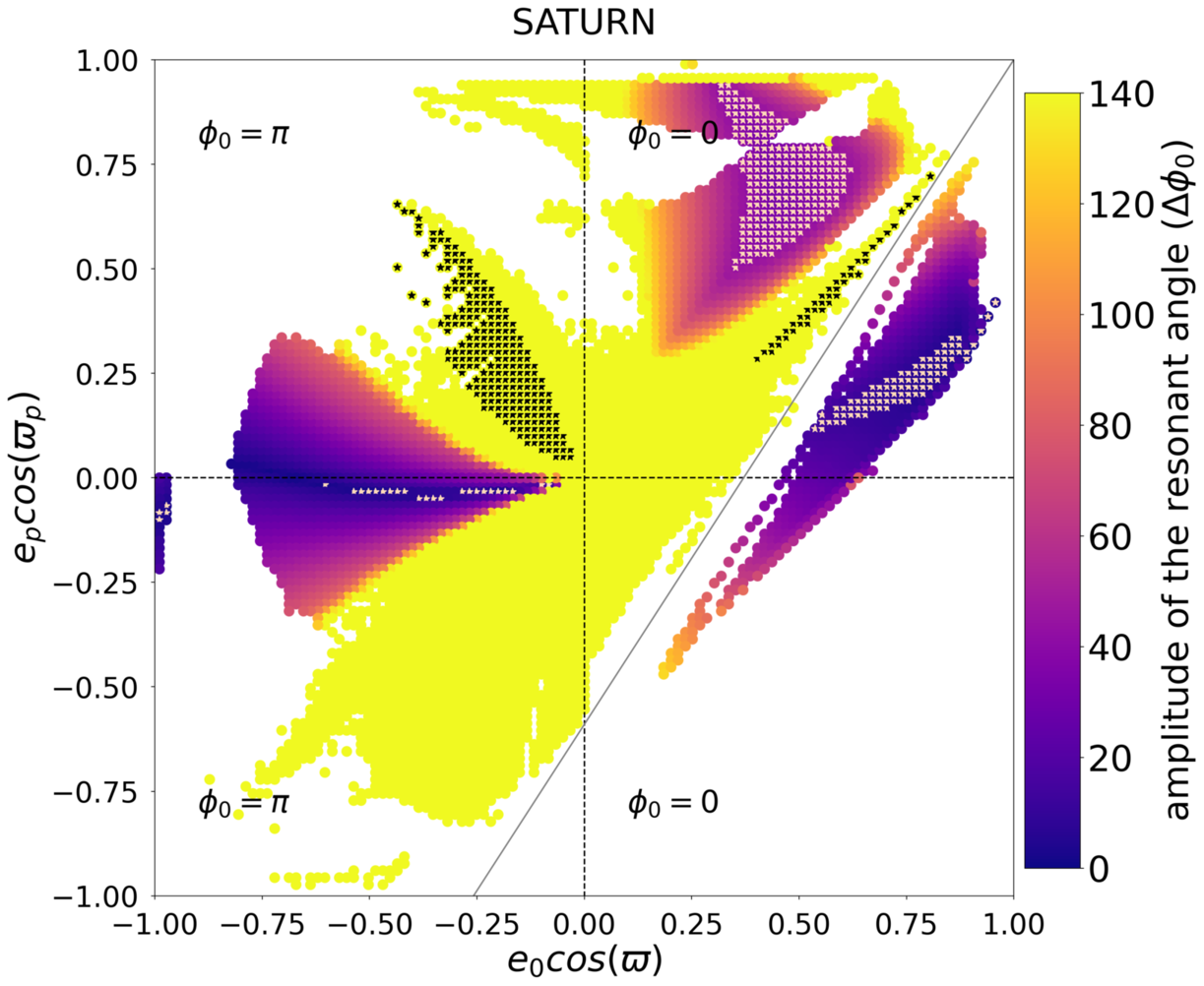}
         \caption{}
         \label{fig12:2sat}
     \end{subfigure}
     
     \caption{Resonant maps for the 1/-2 resonance in the planetary problem when the 2nd planet has Saturns's mass (a) $M=0$; (b) $M=\pi$. The color bar represents the amplitude of the restricted angle ($\phi_{0}$) and the overlaying white symbols indicate the fixed point family where all resonant angles librate around a center. The black symbols indicate libration of $\phi_{3}$ only. The orange and gray lines indicate collision at time zero or after half a period of the external object, respectively.}
     \label{fig:3}
     \end{figure}

The stability maps for the planetary problem when the 2nd planet has Saturn's mass are presented in Figure \ref{fig:3}. The fixed point family of the ER3BP is still present in $Q_1$ while the additional fixed point family within the $\phi_0=0$ libration region which was observed when the 2nd planet has Neptune's mass is displaced to larger $e_p$. There is another fixed point family which appears in $Q_3$ at $e_p\approx0$ within the $\phi_0=\pi$ libration region. In $Q_1$ and $Q_2$ a new family appears which corresponds to libration of $\phi_{3}$ only (represented by black overlaying symbols). 

\begin{figure}
     \centering
     \begin{subfigure}[b]{0.43\textwidth}
         \centering
         \includegraphics[width=\textwidth]{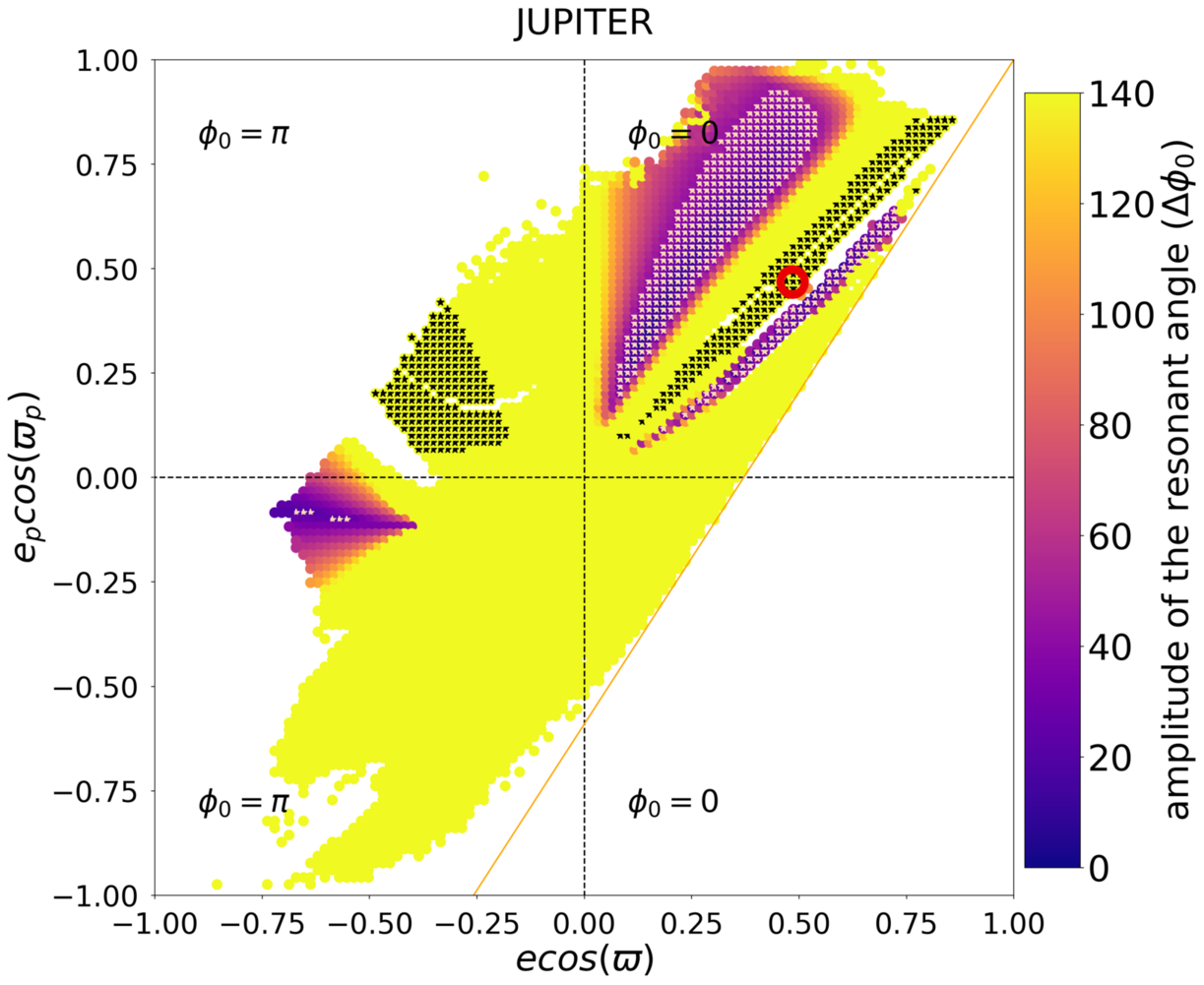}
         \caption{}
         \label{fig12:1jup}
     \end{subfigure}
     \vskip15pt
     \begin{subfigure}[b]{0.43\textwidth}
         \centering
         \includegraphics[width=\textwidth]{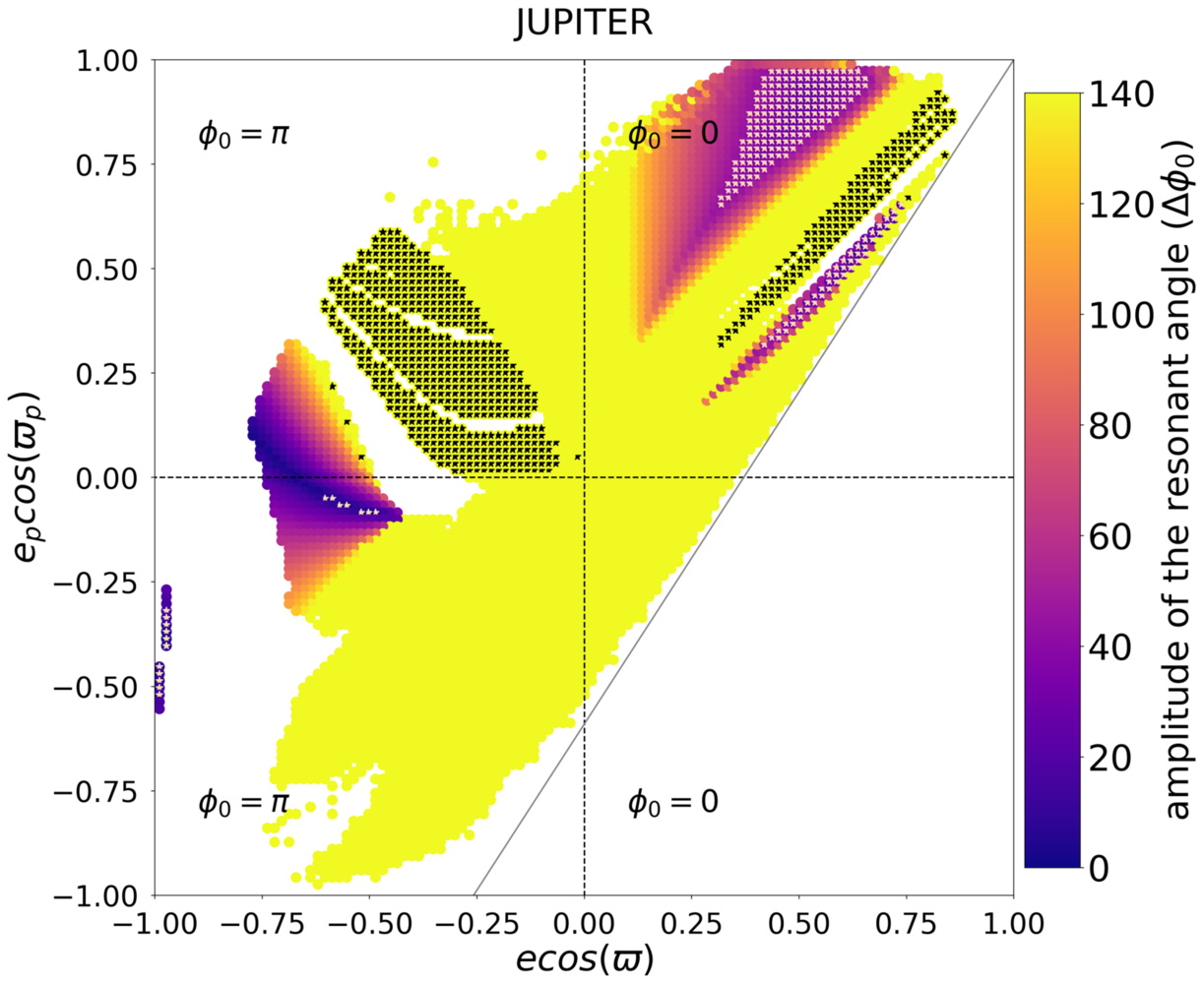}
         \caption{}
         \label{fig12:2jup}
     \end{subfigure}
     
     \caption{Resonant maps for the 1/-2 resonance region considering the third body with Jupiter's mass (a) $M=0$; (b) $M=\pi$. The color bar represents the amplitude of the restricted angle ($\phi_{0}$) and the overlaying points are the fixed point resonant where all resonant angles defined in this region librates around a center. The black symbol represents the libration of $\phi_{3}$. The orange and gray lines indicate collision at time zero or after half a period of the external object, respectively.}
     \label{fig:4}
     \end{figure}

\begin{figure}
    \centering
     \includegraphics[width=0.43\textwidth]{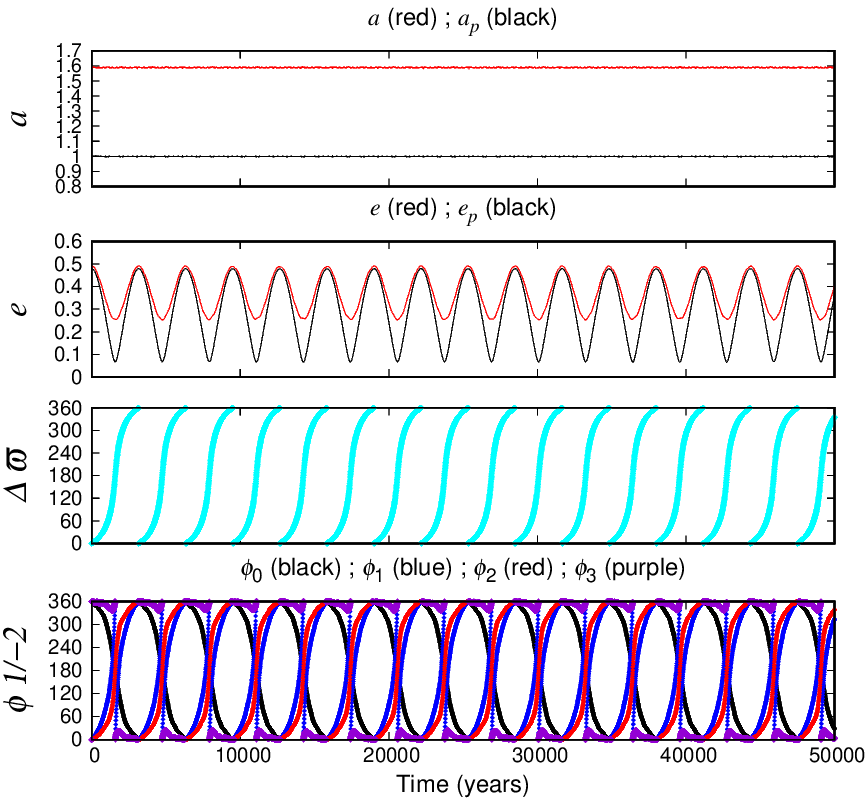}
    \caption{Orbital evolution as a function of time for the initial conditions circled in Figure  \ref{fig12:1jup} ($Q_1$, $M = 0$).  The initial conditions are  $e = 0.49$, $e_p = 0.48$. The 1st panel shows the semi-major axes of both planets, the 2nd panel shows  their eccentricities, the 3rd panel shows the difference $\Delta\varpi$ between the longitudes of pericenter, the 4th panel shows the resonant angles $\phi_0$, $\phi_1$, $\phi_2$, and $\phi_3$.
    }
\label{fig:12jupci}
\end{figure}

The stability maps for the planetary problem when the 2nd planet has
Jupiter's mass are presented in Figure \ref{fig:4} for $M=0$ and $M=\pi$. The resonant families in this case are qualitatively similar to the ones present in Figure \ref{fig:3} but with the difference that libration of $\phi_0$ only is absent in the case of 2 jovian planets. 
In Figure \ref{fig:12jupci} we show the orbital evolution corresponding to the initial conditions marked with red circle in Figure \ref{fig:4} ($e = 0.49$, $e_p = 0.48$). In this case $\phi_3$ is the only angle librating while the others are circulating. The eccentricities of both bodies exhibit simultaneous periodic variations with larger amplitude for the prograde planet.

\subsection{2/-1 Resonance}  

For the retrograde resonance 2/1 the resonant angles analyzed were:

\begin{equation}
    \phi_{0} = -\lambda - 2\lambda_p + 3\varpi
\end{equation}

\begin{equation}
    \phi_1 = -\lambda - 2\lambda_p + 3\varpi_p
\end{equation}

\begin{equation}
   \phi_2 = -\lambda - 2\lambda_p + \varpi_p + 2\varpi
\end{equation}

\begin{equation}
    \phi_3 = -\lambda - 2\lambda_p + 2\varpi_p + \varpi
\end{equation}

Table \ref{tabela2-1sum} presents the summarized results of the 2/-1 resonant configurations, indicating the object's masses and libration angles.

\begin{table*}
\caption{Table reporting the summarized results for the 2/-1 resonance.
The notation $\phi_{0,1,2,3}$ indicates fixed point libration ($\phi_0$ and $\varpi-\varpi_p$ are fixed), and either $\phi_{0}$ or $\phi_3$ indicates libration of a single angle.}
\label{tabela2-1sum}
\centering
\resizebox{0.8\textwidth}{!}{%
\begin{tabular}{cccccc}
\hline
\textbf{Mass} & \multicolumn{4}{c}{Resonance} & \textbf{Figure} \\ \hline
              & $Q_1$  & $Q_2$ & $Q_3$ & $Q_4$ &                 \\ \hline
ER3BP($M=0$) & $\phi_{0,1,2,3}$,$\phi_{0}$ and  $\phi_{3}$ & $\phi_{0,1,2,3}$,$\phi_{0}$ and  $\phi_{3}$ & $\phi_{0,1,2,3}$,$\phi_{0}$ and  $\phi_{3}$ & $\phi_{0,1,2,3}$,$\phi_{0}$ and  $\phi_{3}$ & \ref{fig21:1er3bp} \\
ER3BP($M=\pi$)   & $\phi_{0,1,2,3}$ & - & $\phi_{0,1,2,3}$ & - & \ref{fig21:2er3bp}               \\
NEPTUNE($M=0$)  & $\phi_{0,1,2,3}$,$\phi_{0}$ and  $\phi_{3}$ & $\phi_{0,1,2,3}$,$\phi_{0}$ and  $\phi_{3}$ & $\phi_{0,1,2,3}$,$\phi_{0}$ and  $\phi_{3}$ & $\phi_{0,1,2,3}$,$\phi_{0}$ and  $\phi_{3}$ & \ref{fig21:1nep} \\
NEPTUNE($M=\pi$) & $\phi_{0,1,2,3}$ & - & $\phi_{0,1,2,3}$ & - & \ref{fig21:2nep} \\
SATURN($M=0$)   & $\phi_{0,1,2,3}$ and $\phi_{3}$ & $\phi_{0}$ and $\phi_{3}$ & $\phi_{0,1,2,3}$ and $\phi_{3}$ & $\phi_{0}$ and $\phi_{3}$ & \ref{fig21:1sat} \\
SATURN($M=\pi$)  & $\phi_{0,1,2,3}$ & - & $\phi_{0,1,2,3}$ & - & \ref{fig21:2sat} \\
JUPITER($M=0$)  & $\phi_{0,1,2,3}$ and $\phi_{3}$ & $\phi_{3}$ & $\phi_{0,1,2,3}$ and $\phi_{3}$ & $\phi_{0,1,2,3}$,$\phi_{0}$ and $\phi_{3}$ & \ref{fig21:1jup} \\
JUPITER($M=\pi$) & $\phi_{0,1,2,3}$ & - & $\phi_{0,1,2,3}$ & - & \ref{fig21:2jup} \\ \bottomrule
\end{tabular}%

}
\end{table*}

\begin{figure}
     \centering
     \begin{subfigure}[b]{0.43\textwidth}
         \centering
         \includegraphics[width=\textwidth]{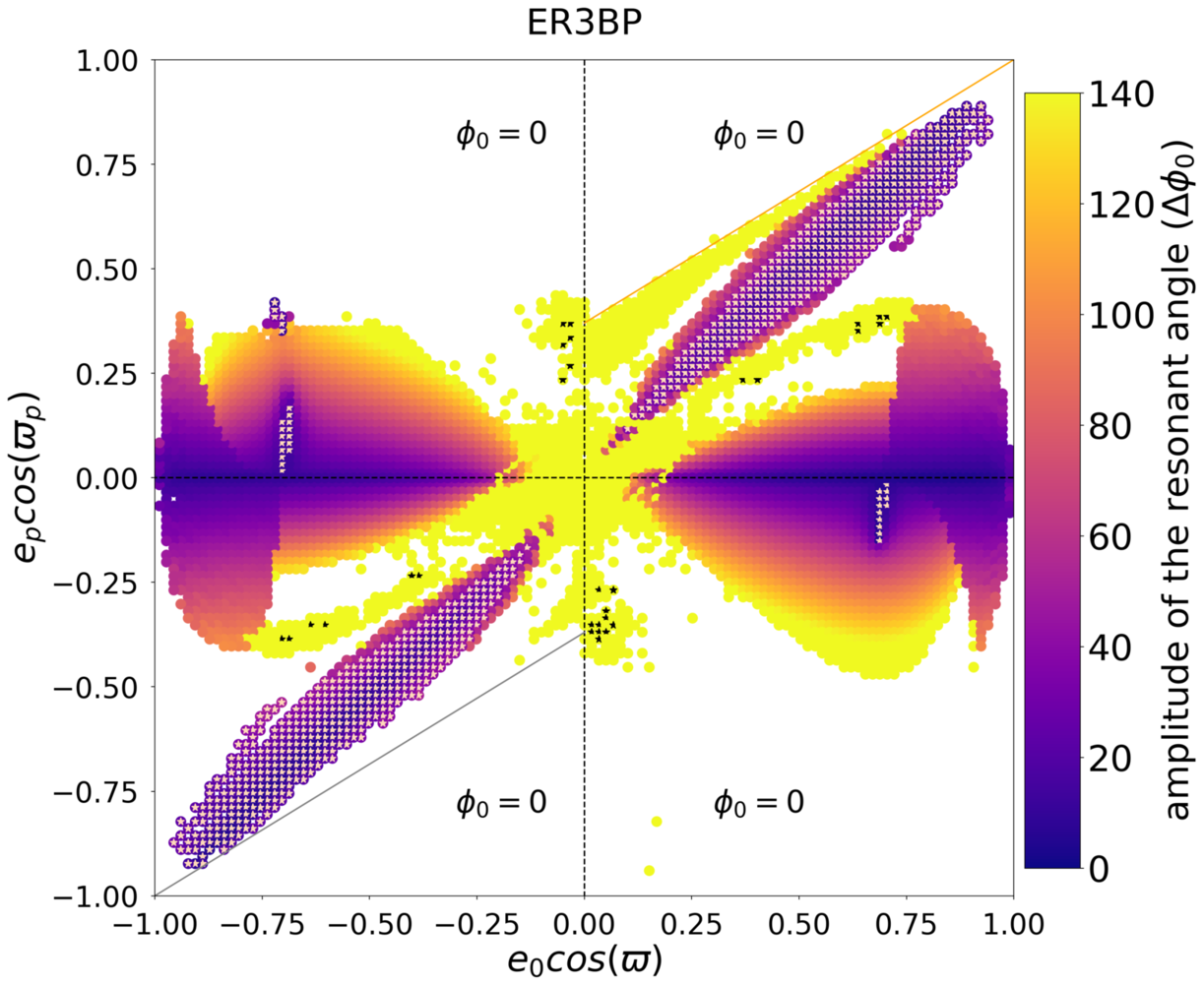}
         \caption{}
         \label{fig21:1er3bp}
     \end{subfigure}
     \vskip15pt
     \begin{subfigure}[b]{0.43\textwidth}
         \centering
         \includegraphics[width=\textwidth]{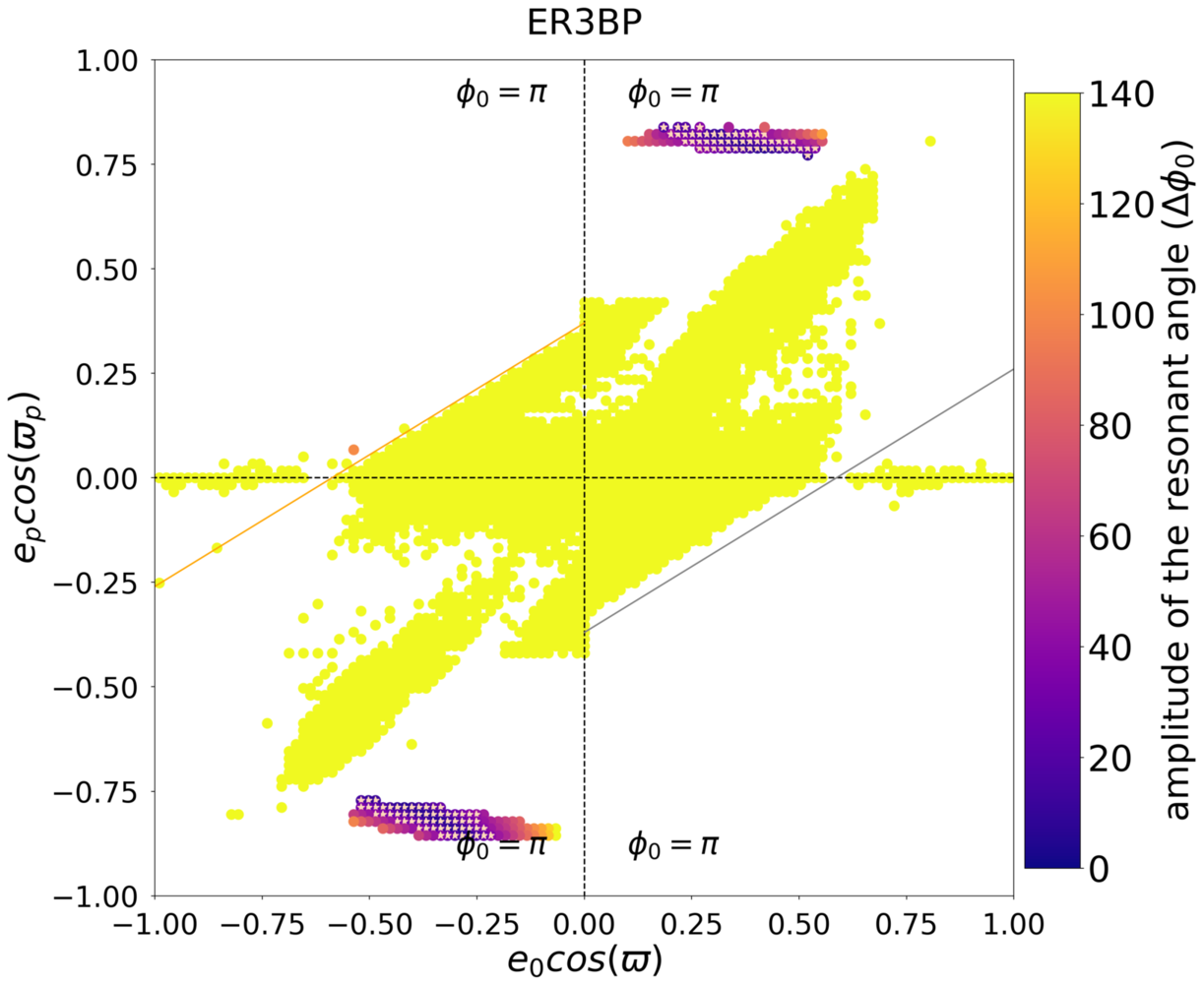}
         \caption{}
         \label{fig21:2er3bp}
     \end{subfigure}
     
     \caption{Resonant maps for the 2/-1 resonance in the elliptic restricted three body problem (a) $M=0$; (b) $M=\pi$. The color bar represents the amplitude of the restricted angle ($\phi_{0}$) and the overlaying white symbols indicate the fixed point family where all resonant angles librate around a center. The black symbol represents the libration of $\phi_{3}$. The orange and gray lines indicate collision at time zero or after half a period of the external object, respectively.}
     \label{fig:5}
     \end{figure}

The stability maps for the ER3BP are presented in Figure \ref{fig:5} for $M = 0$ (a) and   $M=\pi$ (b). The color bar indicates the amplitude of the resonant angle $\phi_{0}$, where dark purple/blue indicates the resonance center. The top panel (a) corresponds to the resonant center $\phi_0 = 0$ while the bottom panel  (b) corresponds to the resonant center $\phi_0 = \pi$. From Figure \ref{fig:configurations} we expect approximate symmetry between the quadrants $Q_1$ and $Q_3$, $Q_2$ and $Q_4$ at fixed $M=0$ or $M=\pi$, which we indeed observe in Figure \ref{fig:5} (a) and (b). There are fixed point families  in the four quadrants for $M=0$:  2 large islands in $Q_1$ and $Q_3$ and 2 smaller islands in $Q_2$ and $Q_4$. There  are regions where  there  is libration of the angle $\phi_0$ near $e_p = 0$ on the left and right of panel (a), and a few points in $Q_1$ and $Q_3$ that correspond to libration of the angle $\phi_3$. For $M=\pi$ there are fixed point families in  quadrants $Q_1$ and $Q_3$.

The 2/-1 resonance at Jupiter to Sun mass ratio in the CR3BP was studied in \cite{morais2016retrograde} and \cite{kotoulas2020planar}. According to these works, the $\phi_0 = 0$ is present  at nearly all values of $e$  in agreement with our results (purple regions at $e_p=0$ on right and left  hand sides of Figure \ref{fig:5} (a)), while the  $\phi_0 = \pi$ exists from $e>0.6$ in the planar problem when $e_p=0$. This latter family does not appear in Figure \ref{fig:5} (b) because it is vertically unstable \citep{kotoulas2020planar}. As we integrate the 3D equations of motion, our problem is not strictly 2D and therefore we do not recover vertically unstable families. In practice, strictly planar systems do not exist hence we do not expect to find real systems in resonant configurations which are vertically unstable. The  2/-1 resonance at Jupiter to Sun mass ratio was also studied in the ER3BP by \cite{kotoulas2020planar}. Our results are in agreement regarding the fixed point point families at in Figure \ref{fig:5} (a). However the  fixed point point families  in Figure \ref{fig:5} (b) have not been reported  by \cite{kotoulas2020planar}.

\begin{figure}
\centering
    \begin{subfigure}[b]{0.43\textwidth}
    \centering
    \includegraphics[width=\textwidth]{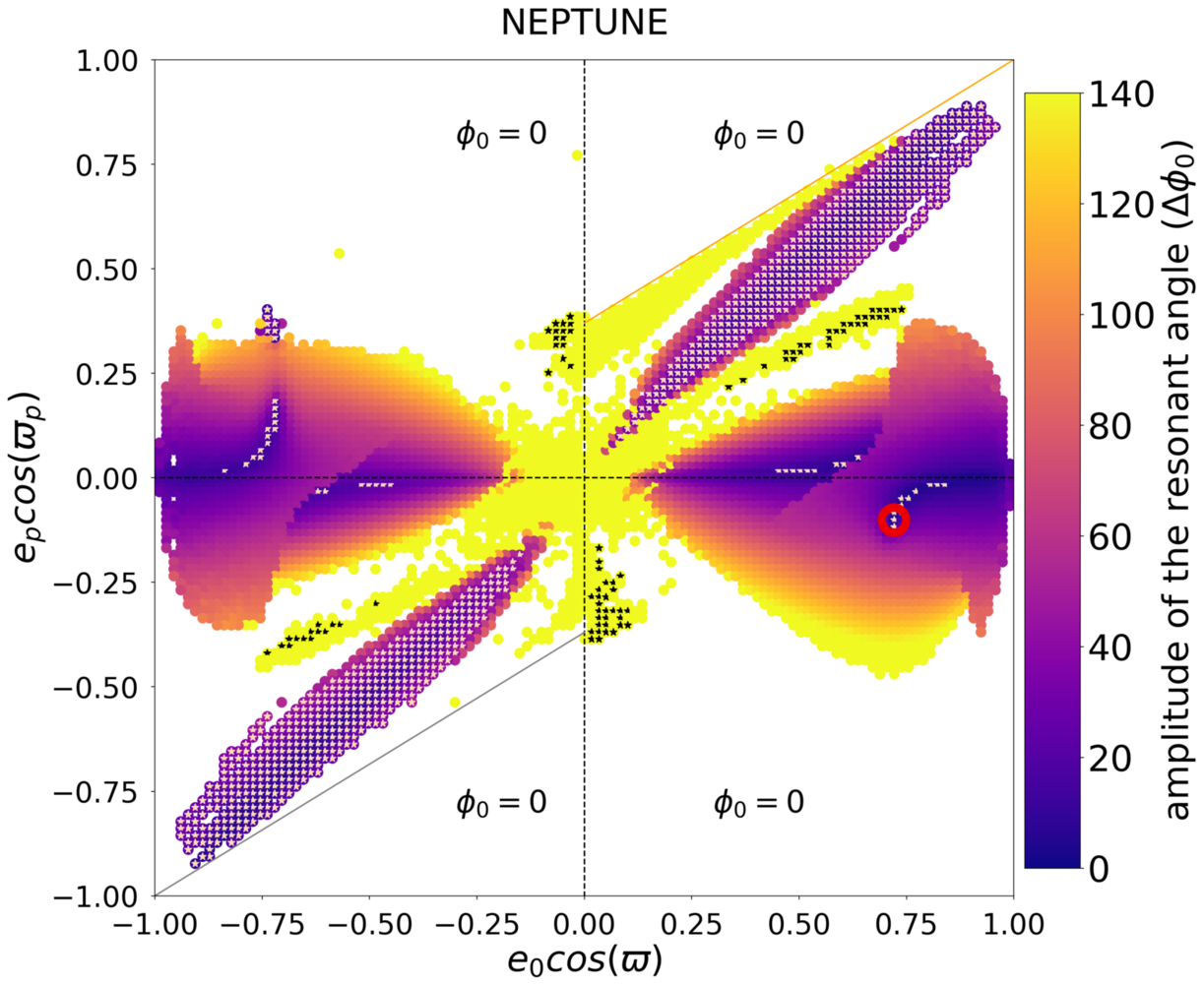}
    \caption{}
    \label{fig21:1nep}
    \end{subfigure}
    \vskip15pt
    \begin{subfigure}[b]{0.43\textwidth}
    \centering
    \includegraphics[width=\textwidth]{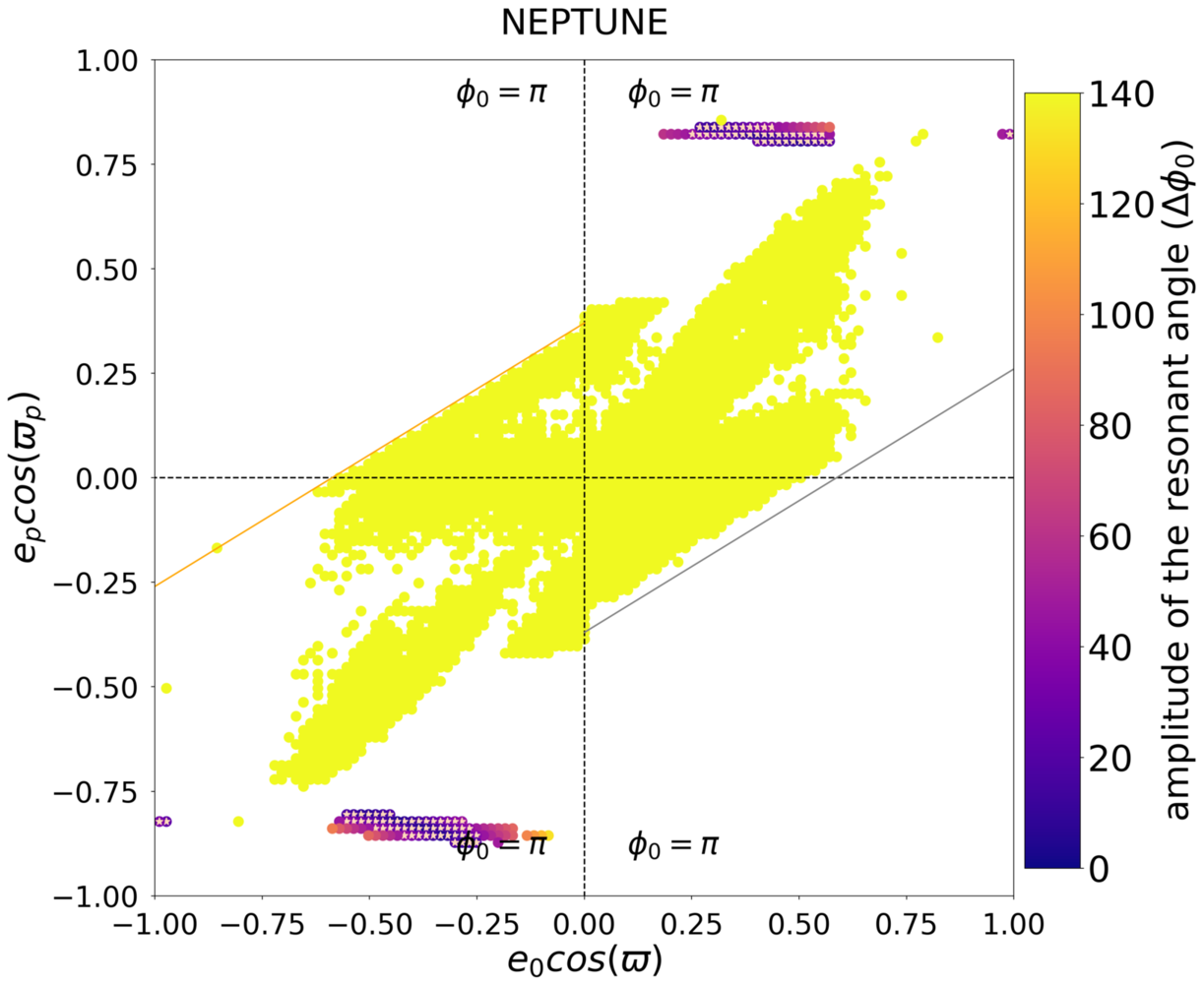}
    \caption{}
    \label{fig21:2nep}
    \end{subfigure}
    \caption{Resonant maps for the 2/-1 resonance in the planetary problem when the 2nd planet has Neptune's mass (a) $M=0$; (b) $M=\pi$. The color bar represents the amplitude of the restricted angle ($\phi_{0}$) and the overlaying white symbols indicate the fixed point family where all resonant angles librate around a center. The black symbol represents the libration of $\phi_{3}$. The orange and gray lines indicate collision at time zero or after half a period of the external object, respectively.}
\label{fig:6}
\end{figure}

\begin{figure}
    \centering
     \includegraphics[width=0.43\textwidth]{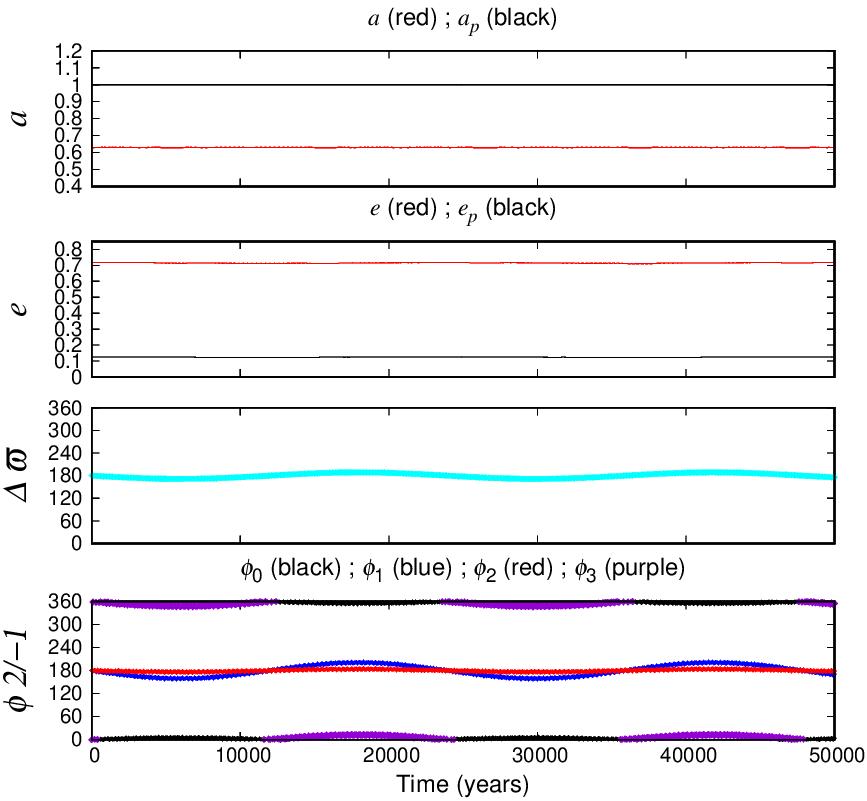}
    \caption{Orbital evolution as a function of time for the initial conditions circled in Figure \ref{fig21:1nep} ($Q_4$, $M=0$).  The initial conditions are  $e = 0.718$, $e_p = 0.125$. The 1st panel shows the semi-major axes of both planets, the 2nd panel shows their eccentricities, the 3rd panel shows the the difference $\Delta\varpi$ between the longitudes of pericenter, the 4th panel shows the resonant angles $\phi_0$, $\phi_1$, $\phi_2$, and $\phi_3$.
    }
\label{fig:21nepp0d180ci}
\end{figure}

The stability maps for the planetary problem when the 2nd planet has Neptune's mass are presented in Figure \ref{fig:6}. At  $e_p\approx 0$ on the left and right hand side of panel (a) there are now 2 fixed point families, corresponding to configurations where the pericenters are aligned (in $Q_1$ and $Q_3$) or anti-aligned (in $Q_2$ and $Q_4$). The latter family was already present in the ER3BP.  The family associated with the 2 large fixed point regions in $Q_1$ and $Q_3$ is nearly identical to the family observed in the ER3BP. There are also 4 regions distributed in the 4 quadrants where the  angle $\phi_2$ librates. It's clear  the symmetry between  $Q_1$ and $Q_2$, and also $Q_3$ and $Q_4$.  For $M = \pi $ we obtain the exactly the same structures that observed in ER3BP: two fixed resonance families at large $e_p$ in $Q_1$ and $Q_3$. In Figure \ref{fig:21nepp0d180ci} we show the orbital evolution corresponding to the initial conditions marked with red circle in Figure \ref{fig:6}. In this case all resonant angles and $\varpi-\varpi_p$ librate around either 0 or $\pi$ with small amplitude as expected for a fixed point resonant family.

\begin{figure}
     \centering
     \begin{subfigure}[b]{0.43\textwidth}
         \centering
         \includegraphics[width=\textwidth]{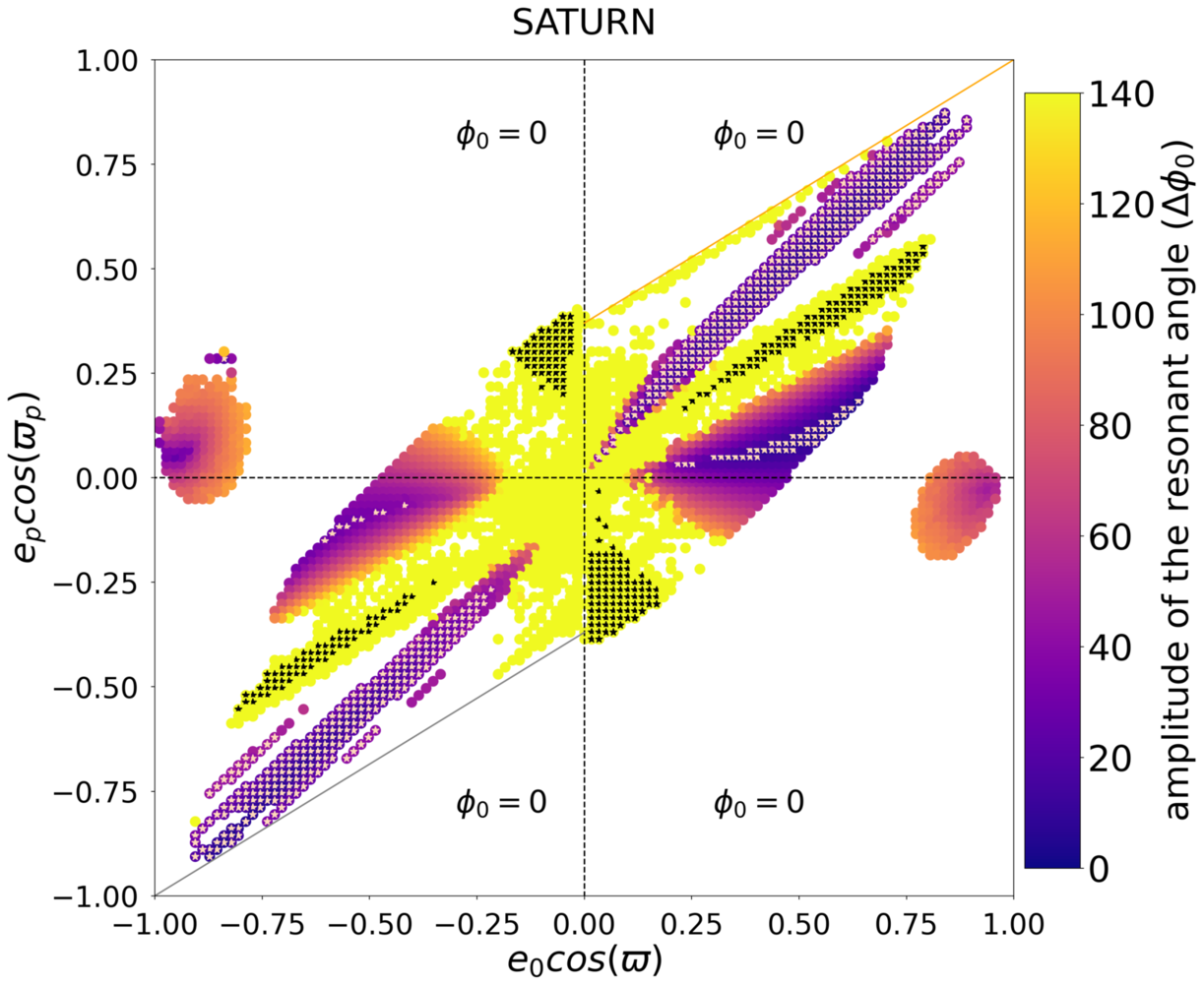}
         \caption{}
         \label{fig21:1sat}
     \end{subfigure}
     \vskip15pt
     \begin{subfigure}[b]{0.43\textwidth}
         \centering
         \includegraphics[width=\textwidth]{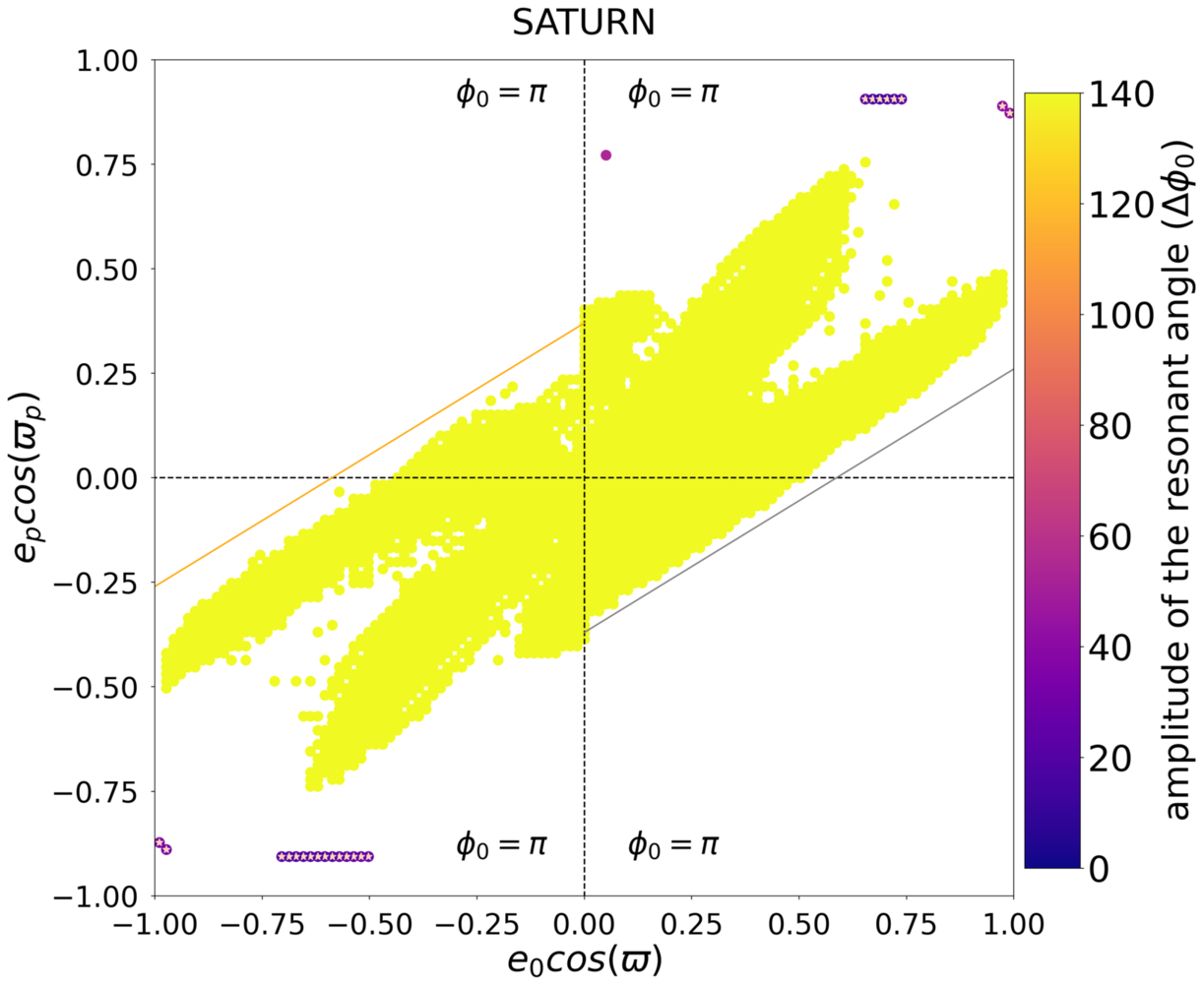}
         \caption{}
         \label{fig21:2sat}
     \end{subfigure}
     
     \caption{Resonant maps for the 2/-1 resonance in the planetary problem when the 2nd planet has Saturns's mass (a) $M=0$; (b) $M=\pi$. The color bar represents the amplitude of the restricted angle ($\phi_{0}$) and the overlaying white symbols indicate the fixed point family where all resonant angles librate around a center. The black symbol represents the libration of $\phi_{3}$. The orange and gray line are the collision lines in $t=0$ and $t=T_{ext}/2$ respectively.}
     \label{fig:7}
     \end{figure}

The stability maps for the planetary problem when the 2nd planet has Saturn's mass are presented in Figure \ref{fig:7}. In Figure \ref{fig:7} (a) ($M=0$) the main difference from the case when the 2nd planet has Neptune's mass is the destruction of the fixed point region near $e_p=0$ at large $e$. In Figure \ref{fig:7} (b) ($M=\pi$) we see that the fixed point family at large $e_p$ in $Q_1$ and $Q_3$ is reduced with respect to the ER3BP and the case of a 2nd planet with Neptune's mass.

\begin{figure}
 \centering
 \begin{subfigure}[b]{0.43\textwidth}
     \centering
     \includegraphics[width=\textwidth]{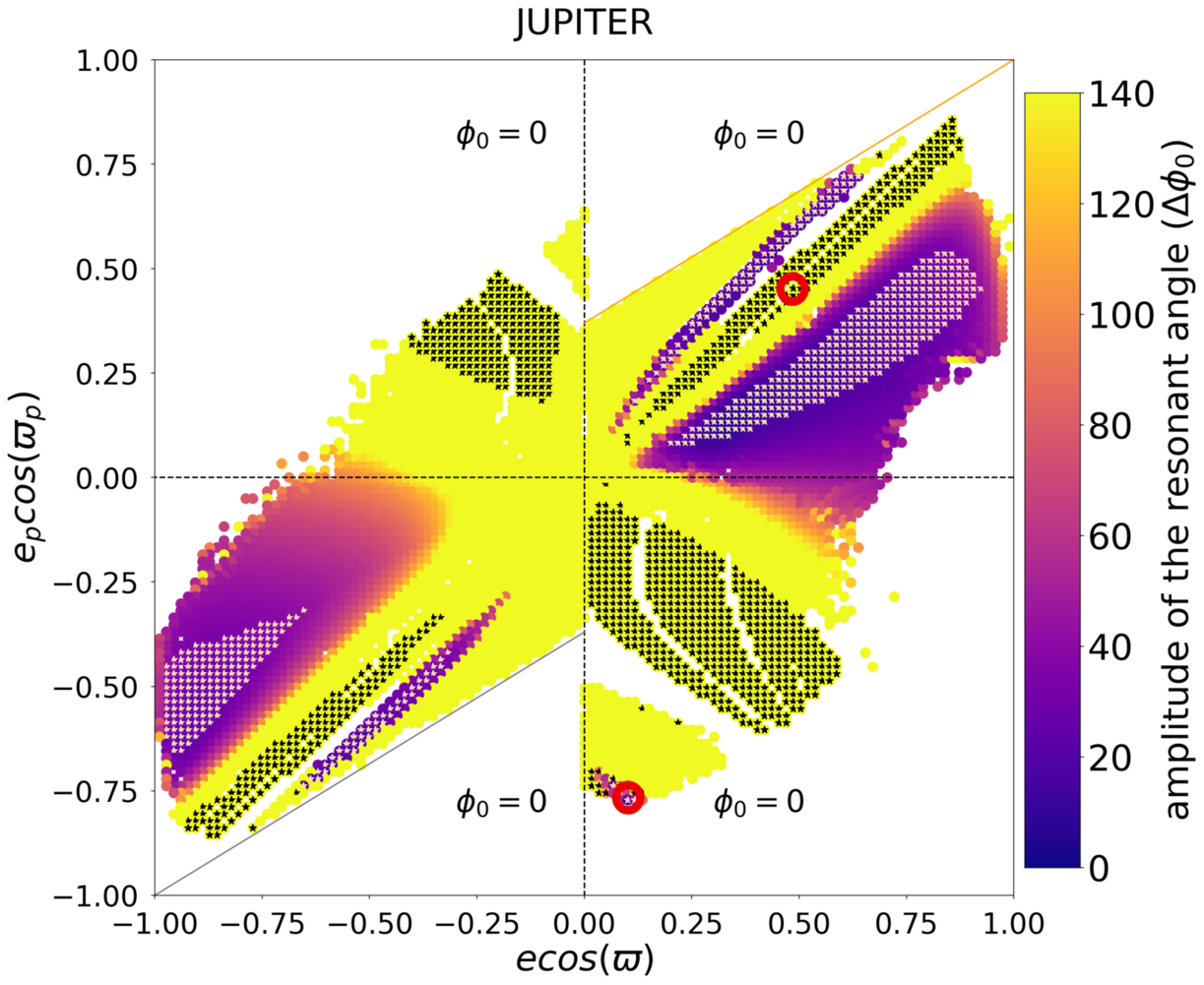}
     \caption{}
     \label{fig21:1jup}
 \end{subfigure}
 \vskip15pt
 \begin{subfigure}[b]{0.43\textwidth}
     \centering
     \includegraphics[width=\textwidth]{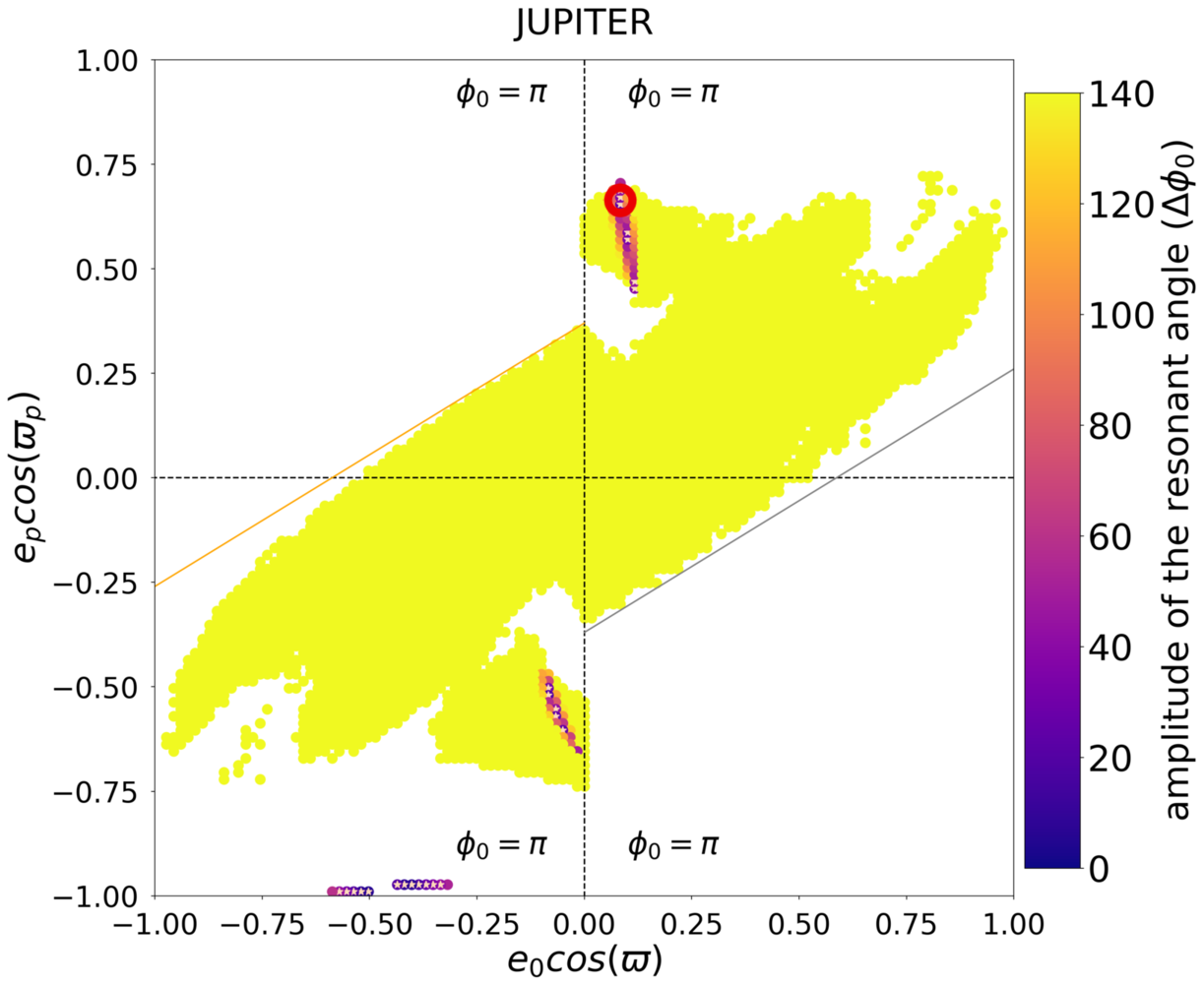}
     \caption{}
     \label{fig21:2jup}
 \end{subfigure}
 
    \caption{Resonant maps for the 2/-1 resonance in the planetary problem when the 2nd planet has Jupiter's mass (a) $M=0$; (b) $M=\pi$. The color bar represents the amplitude of the restricted angle ($\phi_{0}$) and the overlaying white symbols indicate the fixed point family where all resonant angles librate around a center. The black symbol represents the libration of $\phi_{3}$. The orange and gray lines indicate collision at time zero or after half a period of the external object, respectively.}
 \label{fig:8}
 \end{figure}

The stability maps for the planetary problem when the 2nd planet has Jupiter's mass are presented in Figure \ref{fig:8}. In Figure \ref{fig:8} (a) we observe the same structures already described in the case where the 2nd planet has Saturn's mass.  However, although we expect mirroring of the resonant structures observed in $Q_1$ and $Q_3$, and also in $Q_2$ and $Q_4$,  as the initial conditions in these configurations are approximately equivalent with a time lag equal to half a period of the external body (Figure \ref{fig:configurations}), it is now clear that there is no exact symmetry. This is due to the non negligible interaction between the 2 jovian planets during that time lag which implies that the configurations are not exactly equivalent. In particular, there is a small fixed point family in quadrant $Q_4$ near $e_p = 0.75$ which is not present in $Q_2$. In $Q_1$ and $Q_3$ of Figure \ref{fig:8} (b) the fixed point family  present at smaller masses of the 2nd planet does no longer exist but there are 2 new fixed point small islands.

\begin{figure}
    \centering
     \includegraphics[width=0.43\textwidth]{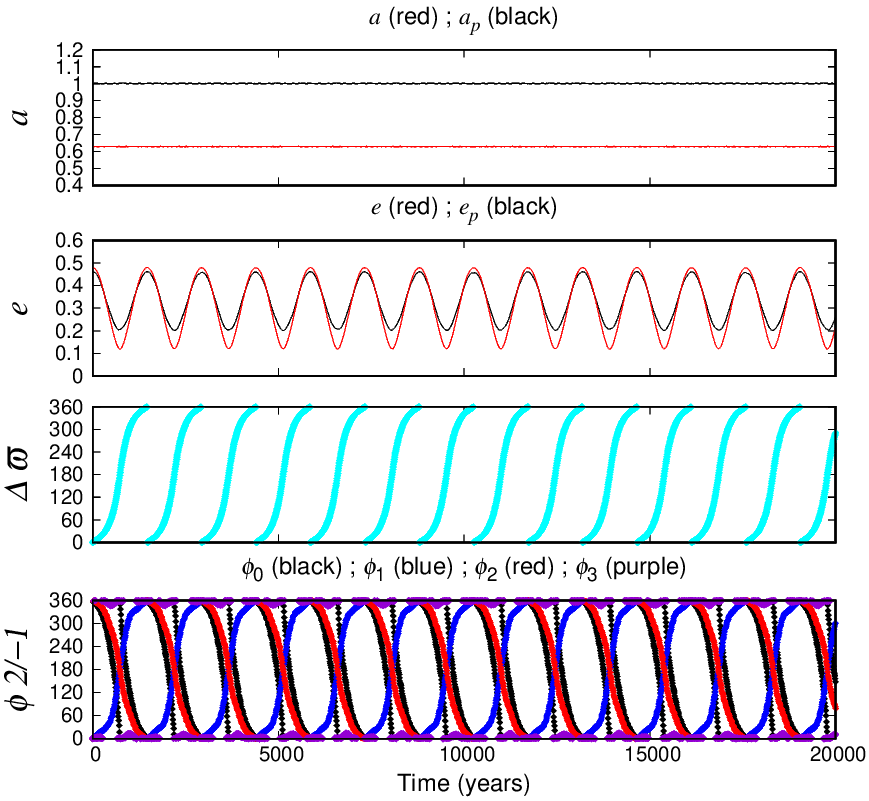}
    \caption{Orbital evolution as a function of time for the initial conditions circled in Figure \ref{fig21:1jup} ($Q_1$, $M=0$).  The initial conditions are  $e = 0.48$, $e_p = 0.46$. The 1st panel shows the semi-major axes of both planets, the 2nd panel shows their eccentricities, the 3rd panel shows the difference $\Delta\varpi$ between the longitudes of pericenter, the 4th panel shows the resonant angles $\phi_0$, $\phi_1$, $\phi_2$, and $\phi_3$.
    }
\label{fig:21jupci}
\end{figure}

\begin{figure}
    \centering
     \includegraphics[width=0.43\textwidth]{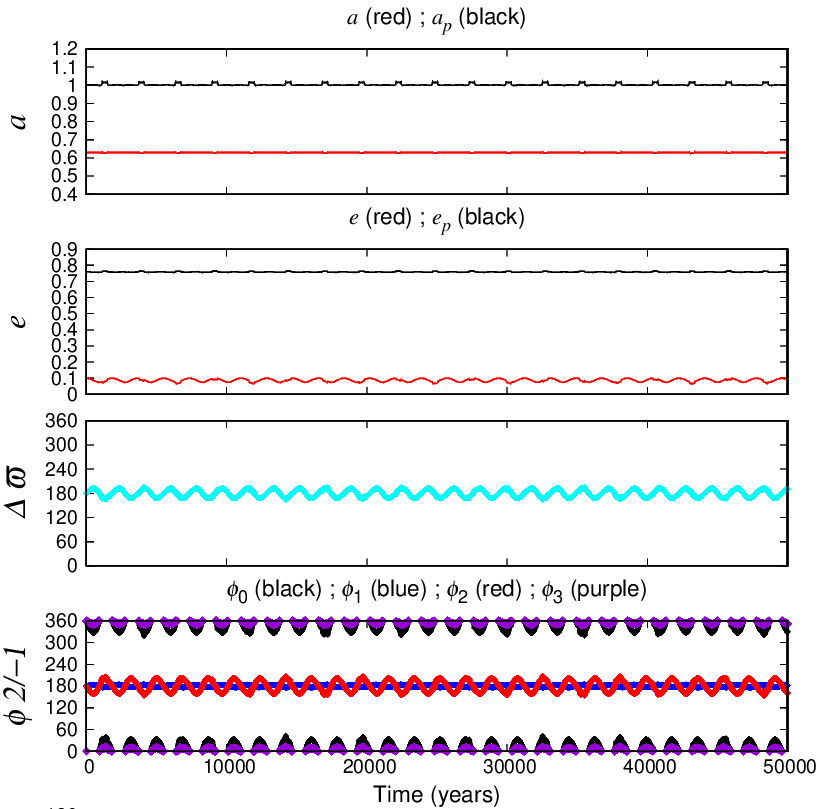}
    \caption{Orbital evolution as a function of time for the initial conditions circled in Figure \ref{fig21:1jup} ($Q_4$, $M = 0$).  The initial conditions are  $e = 0.10$, $e_p = 0.76$. The 1st panel shows the semi-major axes of both planets, the 2nd panel shows their eccentricities, the 3rd panel shows the difference $\Delta\varpi$ between the longitudes of pericenter, the 4th panel shows the resonant angles $\phi_0$, $\phi_1$, $\phi_2$, and $\phi_3$.
    } 
\label{fig:21jupci2}
\end{figure}

\begin{figure}
    \centering
     \includegraphics[width=0.43\textwidth]{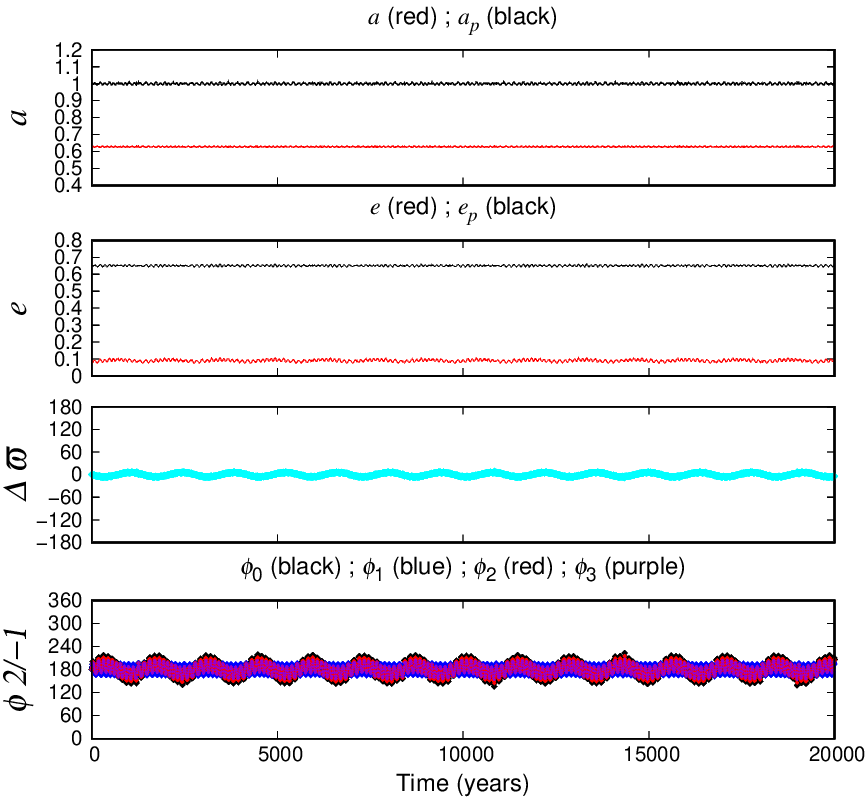}
    \caption{Orbital evolution as a function of time for the initial conditions circled in Figure \ref{fig21:2jup} ($Q_1$, $M = \pi$).  The initial conditions are  $e = 0.08$, $e_p = 0.65$. The 1st panel shows the semi-major axes of both planets, the 2nd panel shows their eccentricities, the 3rd panel shows the difference $\Delta\varpi$ between the longitudes of pericenter, the 4th panel shows the resonant angles $\phi_0$, $\phi_1$, $\phi_2$, and $\phi_3$.
    } 
\label{fig:21jupci3}
\end{figure}

In Figure \ref{fig:21jupci} we show the orbital evolution corresponding to the initial condition marked with red circle in $Q_1$ of Figure \ref{fig:8} (a). In this case only the resonant angle $\phi_2$ librates while all other resonant angles circulate. Unlike the behavior observe in $1/-2$ resonance, in $2/-1$ resonance the amplitude of the eccentricity of the retrograde planet is higher than the prograde body. In Figure \ref{fig:21jupci2}, we show the orbital evolution corresponding to the initial condition marked with red circle in $Q_4$ of Figure \ref{fig:8} (a). In this case there is a fixed point resonance where $\phi_1$ and $\phi_3$ librate around 0 while $\phi_0$ and $\phi_2$ around $\pi$. Figure \ref{fig:21jupci3} shows that the initial condition marked with red circle in $Q_1$ of Figure \ref{fig:8} (b), corresponds to a fixed point resonance.

\subsection{1/-1 resonance} 

For resonance 1/-1 the resonant angles analyzed were:

\begin{equation}
    \phi_{0} = -\lambda - \lambda_p + 2\varpi
\end{equation}

\begin{equation}
    \phi_{1} = -\lambda - \lambda_p + 2\varpi_p
\end{equation}

\begin{equation}
   \phi_{2} = -\lambda - \lambda_p + \varpi_p + \varpi
\end{equation}

Table \ref{tabela1-1sum} presents the summarized results of the 1/-1 resonant configurations, indicating the object's masses and libration angles. 

\begin{table*}
\centering
\caption{Table reporting the summarized results for the 1/-1 resonance.
The notation $\phi_{0,1,2}$ indicates fixed point libration ($\phi_0$ and $\varpi-\varpi_p$ are fixed), while $\phi_{0}$  indicates libration of the CR3BP angle.}
\label{tabela1-1sum}
\resizebox{0.7\textwidth}{!}{%
\begin{tabular}{cccccc}
\hline
\textbf{Mass} & \multicolumn{4}{c}{Resonance} & \textbf{Figure} \\ \hline
              & $Q_1$  & $Q_2$ & $Q_3$ & $Q_4$ &                 \\ \hline
ER3BP($M=0$)    & $\phi_{0}$ & $\phi_{0}$ & $\phi_{0}$ & $\phi_{0}$ &\ref{fig11:1er3bp}\\
ER3BP($M=\pi$)   & $\phi_{0,1,2}$ and $\phi_{0}$ & $\phi_{0}$ & $\phi_{0}$ & $\phi_{0,1,2}$ and $\phi_{0}$ &\ref{fig11:2er3bp}\\
NEPTUNE($M=0$)  & $\phi_{0,1,2}$ and $\phi_{0}$ & $\phi_{0}$ & $\phi_{0}$ & $\phi_{0}$ &\ref{fig11:1nep}\\
NEPTUNE($M=\pi$) & $\phi_{0,1,2}$ and $\phi_{0}$ & $\phi_{0}$ & $\phi_{0,1,2}$ & $\phi_{0,1,2}$ and $\phi_{0}$ &\ref{fig11:2nep}\\
SATURN($M=0$)   & $\phi_{0}$ & $\phi_{0}$ & $\phi_{0}$ & - &\ref{fig11:1sat}\\
SATURN($M=\pi$)  & $\phi_{0,1,2}$ & $\phi_{0}$ & $\phi_{0}$ & $\phi_{0}$ &\ref{fig11:2sat}\\
JUPITER($M=0$)  & - & - & $\phi_{0,1,2}$ & - &\ref{fig11:1jup}\\
JUPITER($M=\pi$) & $\phi_{0,1,2}$ & - & - & - &\ref{fig11:2jup}\\ \bottomrule
\end{tabular}%

}

\end{table*}

\begin{figure}
     \centering
     \begin{subfigure}[b]{0.43\textwidth}
         \centering
         \includegraphics[width=\textwidth]{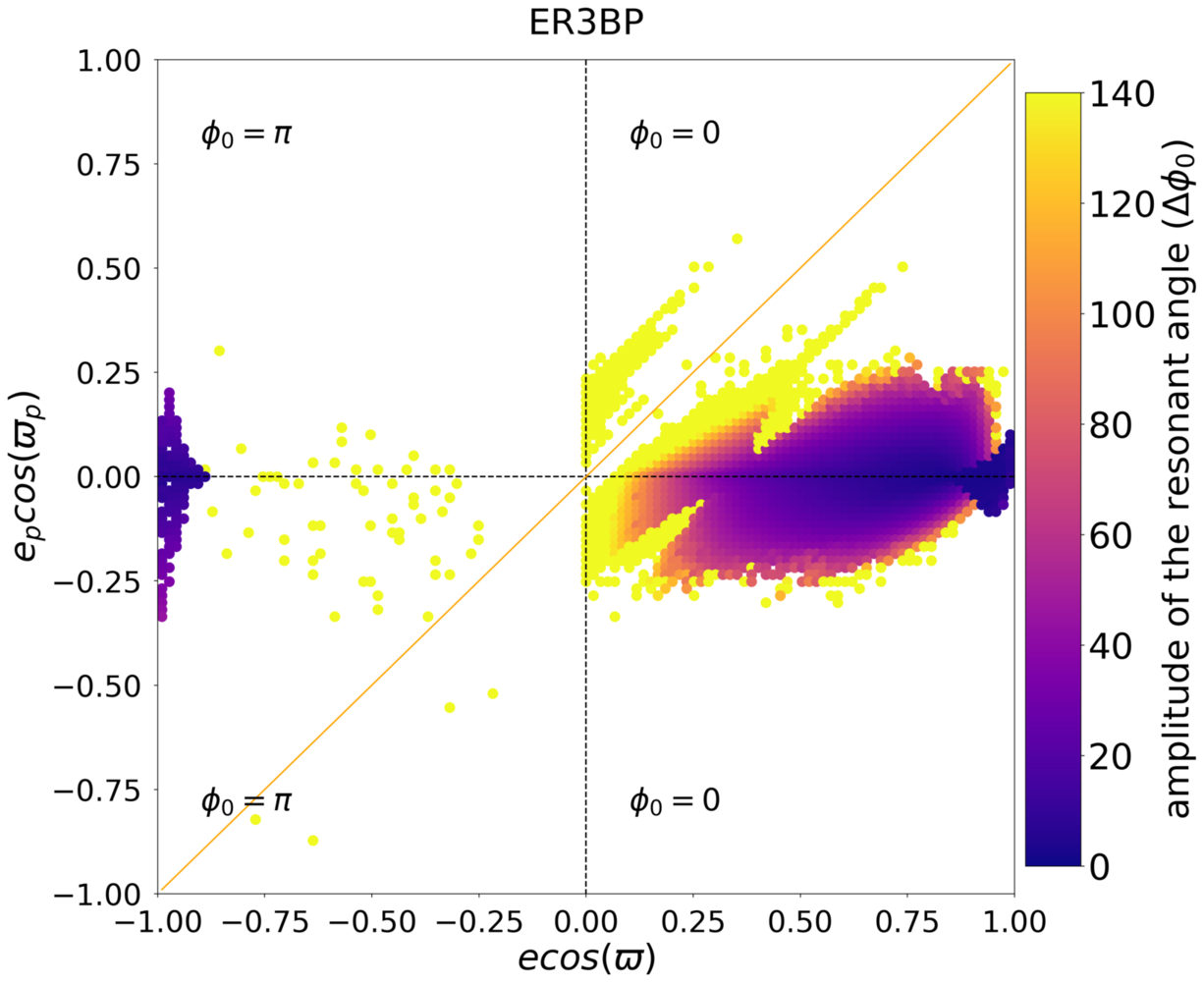}
         \caption{}
         \label{fig11:1er3bp}
     \end{subfigure}
     \vskip15pt
     \begin{subfigure}[b]{0.43\textwidth}
         \centering
         \includegraphics[width=\textwidth]{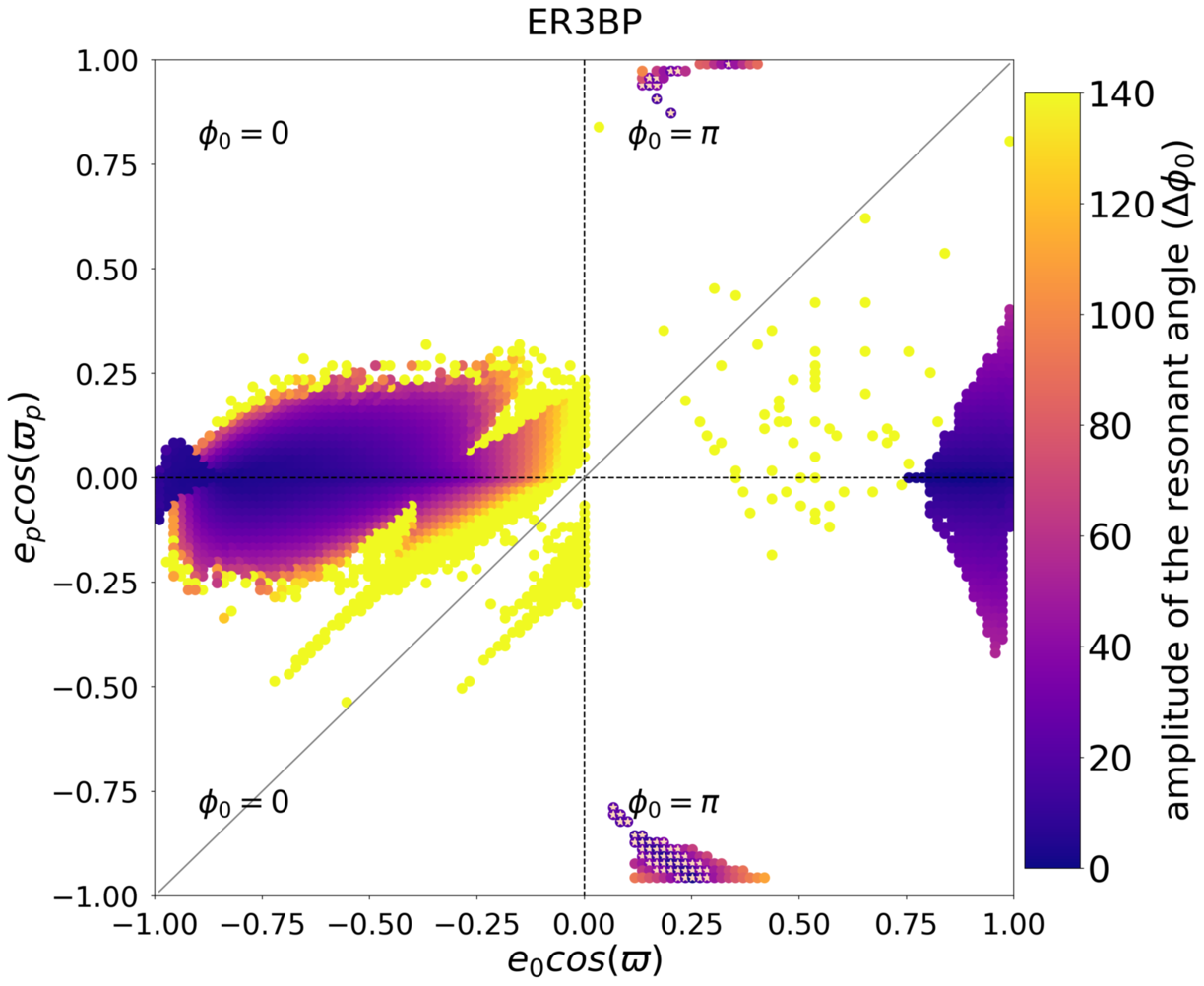}
         \caption{}
         \label{fig11:2er3bp}
     \end{subfigure}
     
     \caption{Resonant maps for the 1/-1 resonance in the elliptic restricted three body problem (a) $M=0$; (b) $M=\pi$. The color bar represents the amplitude of the restricted angle ($\phi_{0}$) and the overlaying white symbols indicate the fixed point family where all resonant angles librate around a center. The orange and gray lines indicate collision at time zero or after half a period of the external object, respectively.}
     \label{fig:9}
\end{figure}

The stability maps for the ER3BP are presented in Figure \ref{fig:9} for $M = 0$ (a) and   $M=\pi$ (b). The color bar indicates the amplitude of the resonant angle $\phi_{0}$, where dark purple/blue indicates the resonance center. The top panel (a) corresponds to  $M = 0$, which implies $\phi_0=0$ in $Q_1$, $Q_4$, and $\phi_0=\pi$ in $Q_2$, $Q_3$, while the bottom panel  (b) corresponds to $M = \pi$, which implies $\phi_0=\pi$ in $Q_1$, $Q_4$, and $\phi_0=0$ in $Q_2$, $Q_3$.  From Figure \ref{fig:configurations} we expect approximate symmetry between $Q_1$ ($Q_3$) at $M=0$ and $Q_3$ ($Q_1$) at $M=\pi$, and also between $Q_2$ ($Q_4$) at $M=0$ and $Q_4$ ($Q_2$) at $M=\pi$. However, it is again clear that this symmetry is not exact.  In this case we did not observe fixed point families for $M=0$. We have obtained fixed families for $M=\pi$ in $Q_1$ and $Q_4$, however these families occur when the planet's eccentricity, $e_p$, is near 1 and thus their long term stability requires further investigation. Moreover, the region corresponding to $\phi_0=\pi$ libration at large eccentricity $e$, is wider when $M=\pi$ (Figure \ref{fig:9} (b)) than when $M=0$ (Figure \ref{fig:9} (a)).

\begin{figure}
     \centering
     \begin{subfigure}[b]{0.43\textwidth}
         \centering
         \includegraphics[width=\textwidth]{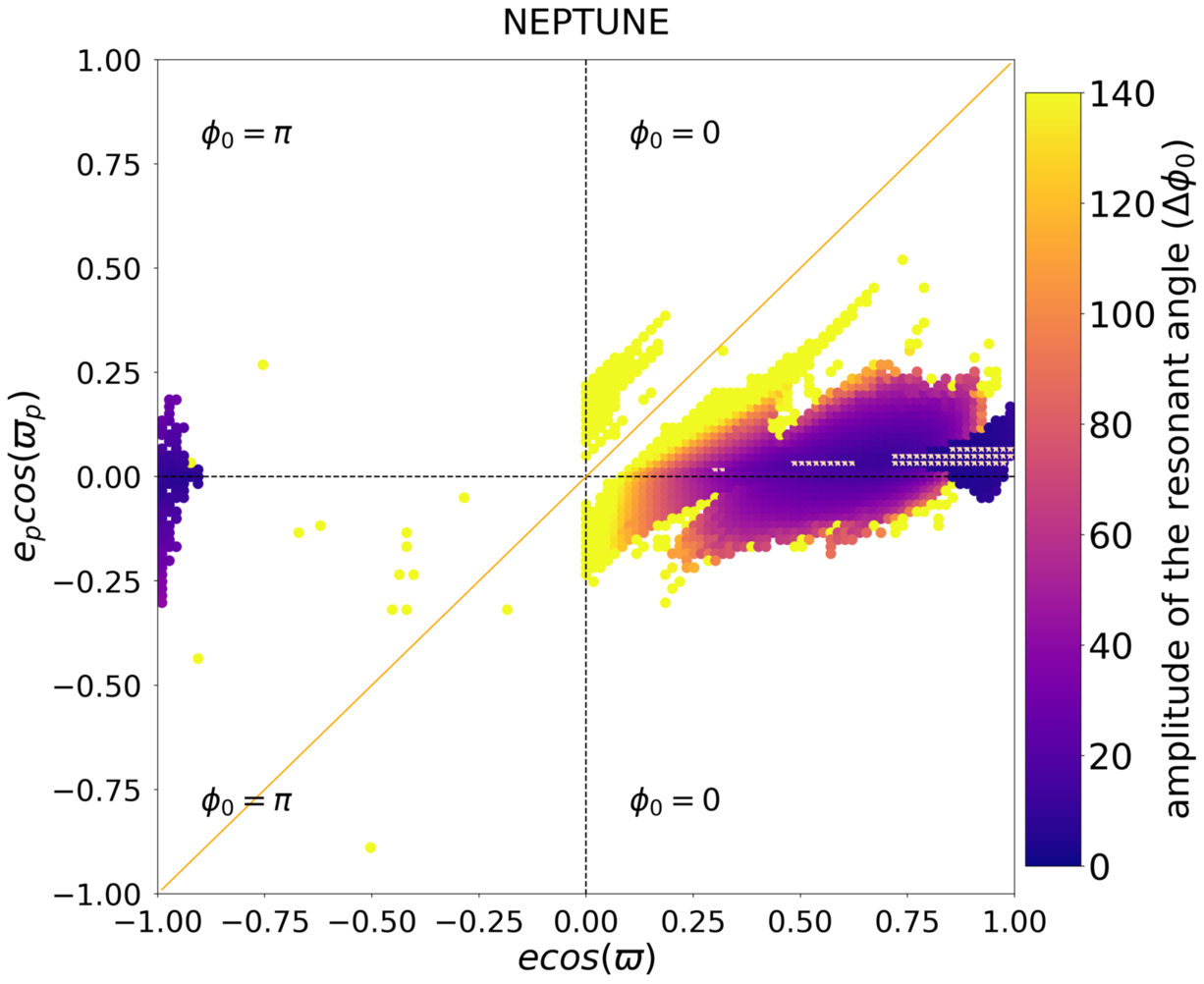}
         \caption{}
         \label{fig11:1nep}
     \end{subfigure}
     \vskip15pt
     \begin{subfigure}[b]{0.43\textwidth}
         \centering
         \includegraphics[width=\textwidth]{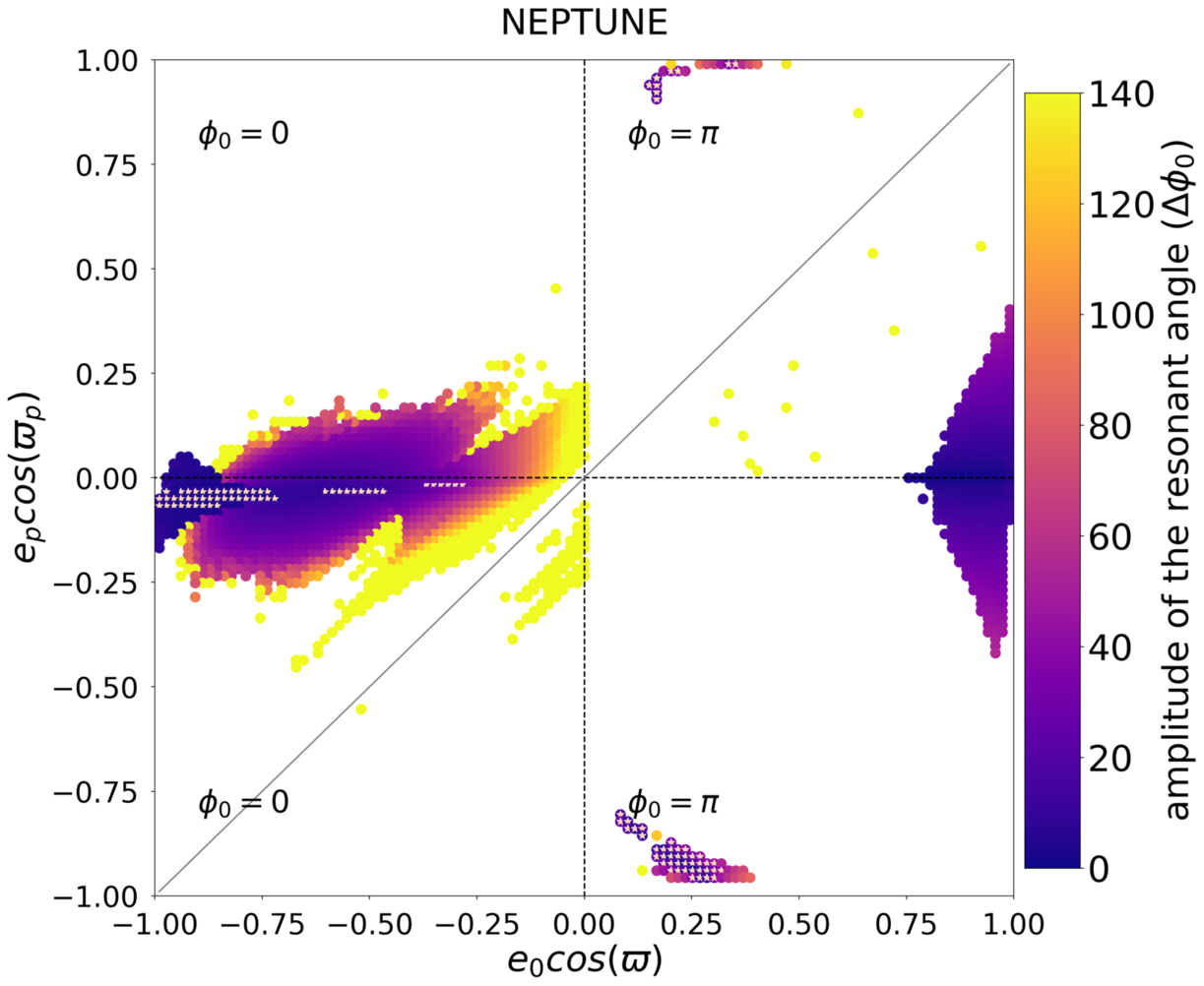}
         \caption{}
         \label{fig11:2nep}
     \end{subfigure}
     
     \caption{Resonant maps for the 1/-1 resonance in the planetary problem when the 2nd planet has Neptune's mass (a) $M=0$; (b) $M=\pi$. The color bar represents the amplitude of the restricted angle ($\phi_{0}$) and the overlaying white symbols indicate the fixed point family where all resonant angles librate around a center. The orange and gray lines indicate collision at time zero or after half a period of the external object, respectively.}
     \label{fig:10}
\end{figure}

The 1/-1 resonance at Jupiter to Sun mass ratio in the CR3BP was studied in \cite{morais2016numerical,morais2019periodic}. There are 2 planar resonant modes: $\phi_0 = 0$ which is stable when $e>0.1$ and whose center occurs at the nominal resonance location, $a=1$, when $e>0.5$  in agreement with our results (purple regions at $e_p=0$ on right  hand side of Figure \ref{fig:9} (a) and left hand side of  Figure \ref{fig:9} (b));  $\phi_0 = \pi$  which occurs at the nominal resonant location $a=1$ when $e>0.75$ also  in agreement with our results (purple regions at $e_p=0$ on left  hand side of Figure \ref{fig:9} (a) and right hand side of  Figure \ref{fig:9} (b)).

The stability maps for the planetary problem when the 2nd planet has Neptune's mass are presented in Figure \ref{fig:10}. At  the center of $\phi_0$ family of the ER3BP there is now a fixed point family near $e_p=0$ seen on right side of Figure \ref{fig:10} (a) and left side of Figure \ref{fig:10} (a). Although this fixed point family appears to have gaps in Figure \ref{fig:10}, a zoom of this region with a higher resolution grid shows that it is in fact a continuous family.

\begin{figure}
     \centering
     \begin{subfigure}[b]{0.43\textwidth}
         \centering
         \includegraphics[width=\textwidth]{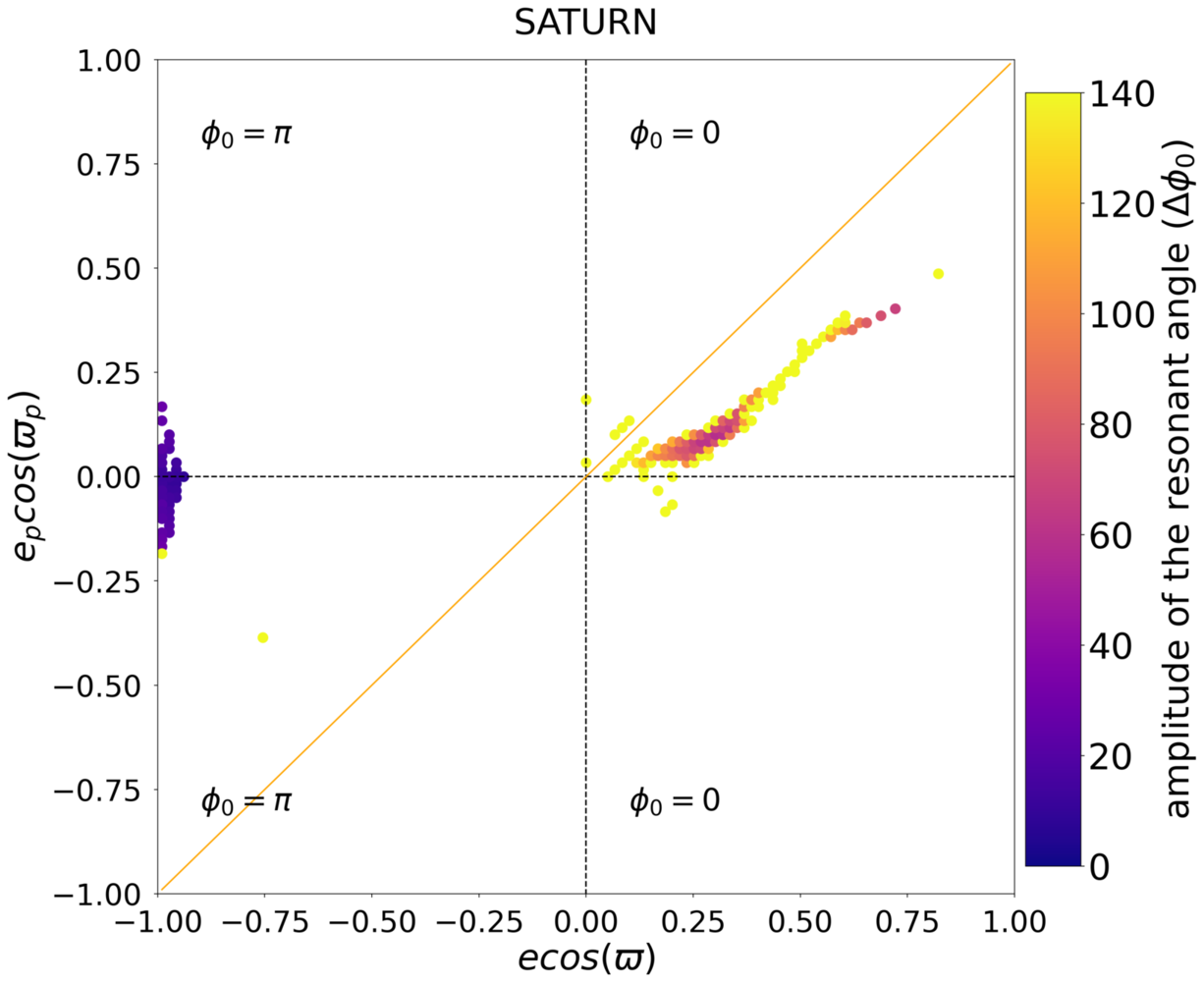}
         \caption{}
         \label{fig11:1sat}
     \end{subfigure}
     \vskip15pt
     \begin{subfigure}[b]{0.43\textwidth}
         \centering
         \includegraphics[width=\textwidth]{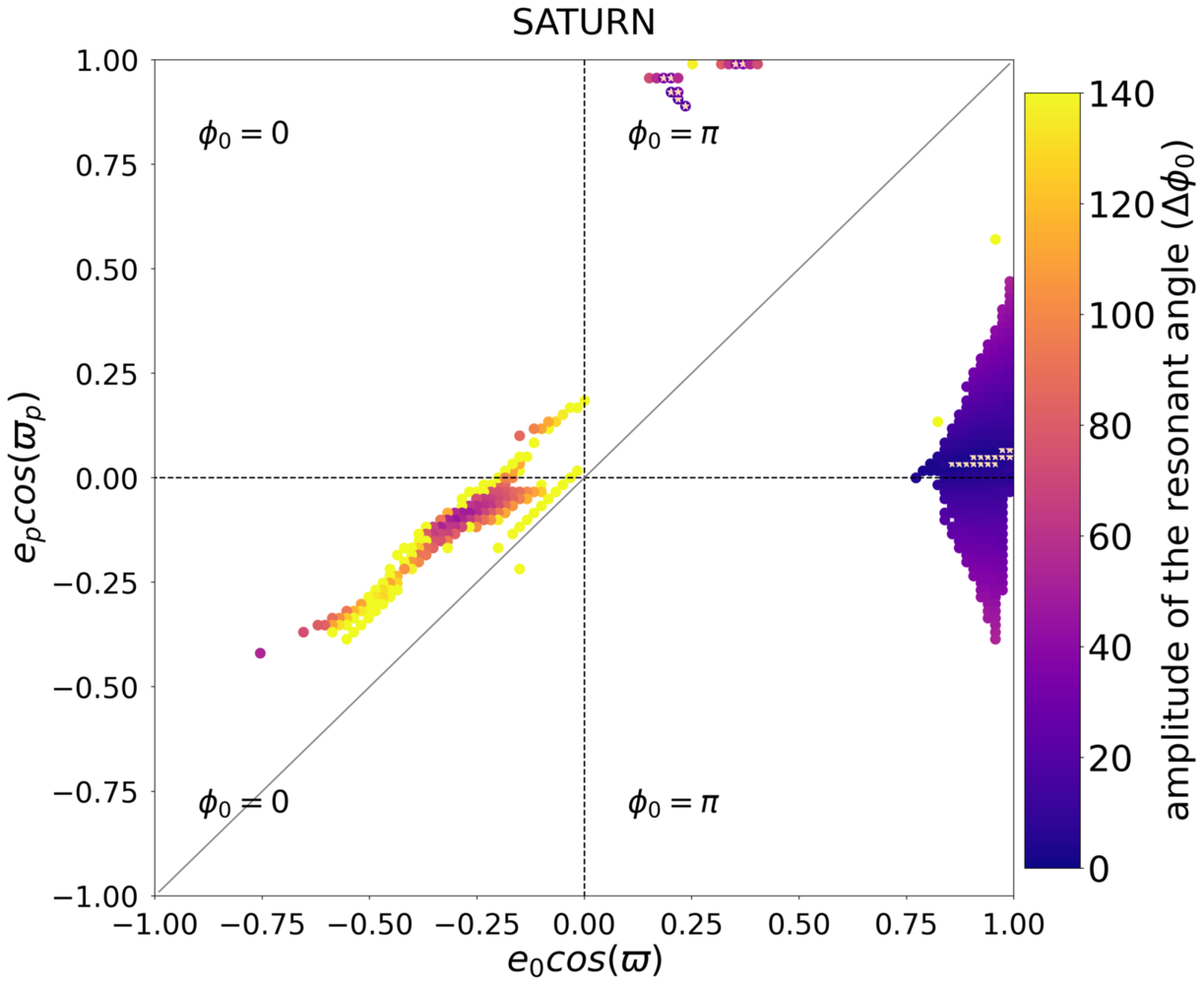}
         \caption{}
         \label{fig11:2sat}
     \end{subfigure}
     
     \caption{Resonant maps for the 1/-1 resonance in the planetary problem when the 2nd planet has Saturn's mass (a) $M=0$; (b) $M=\pi$. The color bar represents the amplitude of the restricted angle ($\phi_{0}$) and the overlaying white symbols indicate the fixed point family where all resonant angles librate around a center. The orange and gray lines indicate collision at time zero or after half a period of the external object, respectively.}
     \label{fig:11}
\end{figure}

\begin{figure}
         \centering
         \includegraphics[width=0.48\textwidth]{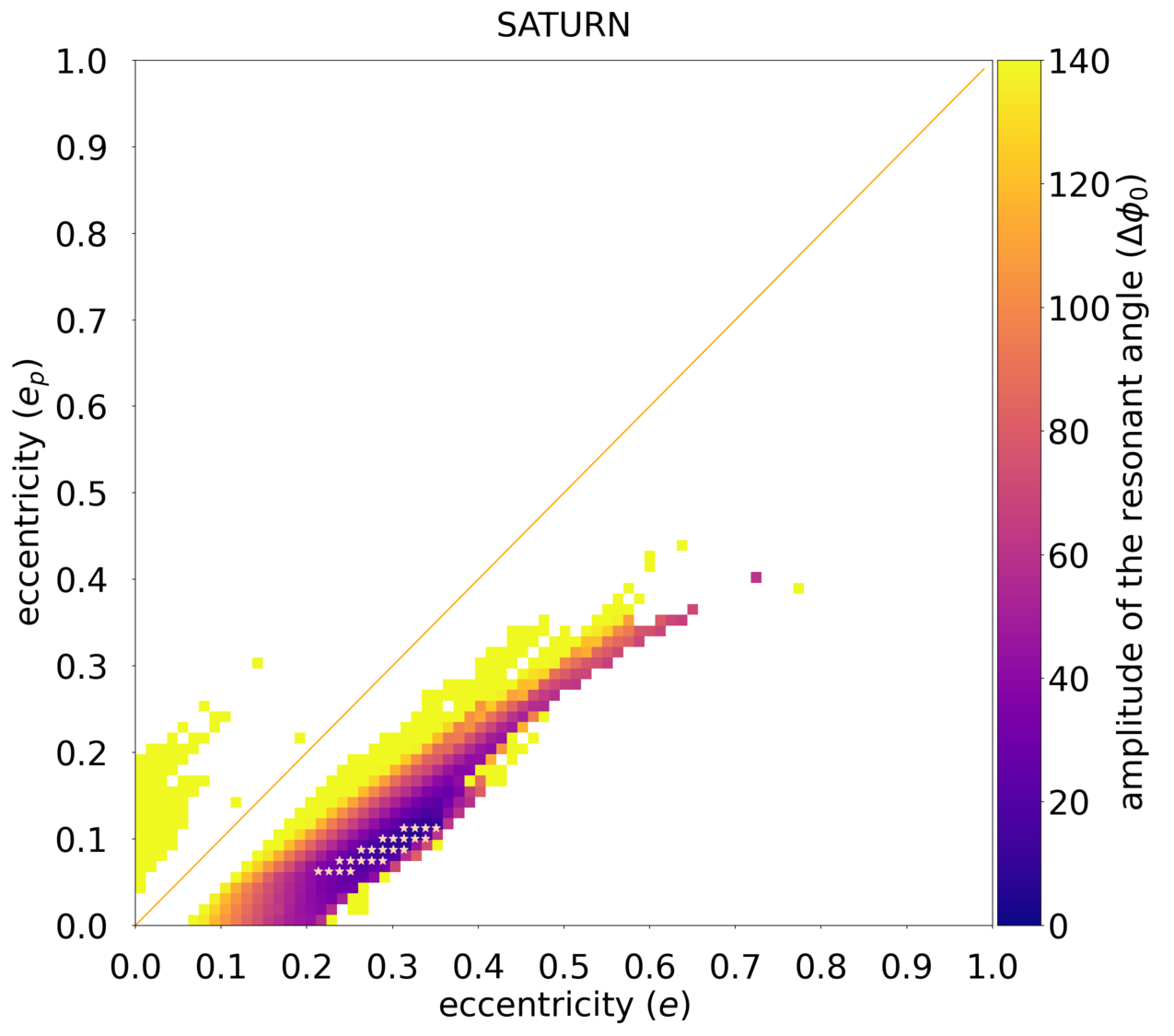}
         \caption{Resonant map for the 1/-1 resonance in the planetary problem when the 2nd planet has Saturns's mass with initial conditions and $a=1.01$, $M=0$, $\varpi=\varpi_p=0$. The color bar represents the amplitude of the restricted angle ($\phi_{0}$) and the overlaying white symbols indicate the fixed point family where all resonant angles librate around a center.}
         \label{fig:6b}
\end{figure}

The stability maps for the planetary problem when the 2nd planet has Saturn's mass are presented in Figure \ref{fig:11}. At the nominal resonance location $a=1$ we only obtain stable solutions with $\phi_0=\pi$ at large $e$. When $e_p\approx 0$ there is a fixed point family at the center of the $\phi_0=\pi$ region which appears with starting $M=\pi$ (Figure \ref{fig:11} (b)) but not $M=0$ (Figure \ref{fig:11} (a)). 
Figure \ref{fig:6b} shows the stability map in $Q_1$ for the planetary problem when the 2nd planet has Saturn's mass, with initial conditions $a=1.01$, $M=0$, $\varpi=\varpi_p=0$ (hence $\phi_0=0$). The fixed point family in this case is displaced  from the nominal resonance location.

\begin{figure}
     \centering
     \begin{subfigure}[b]{0.43\textwidth}
         \centering
         \includegraphics[width=\textwidth]{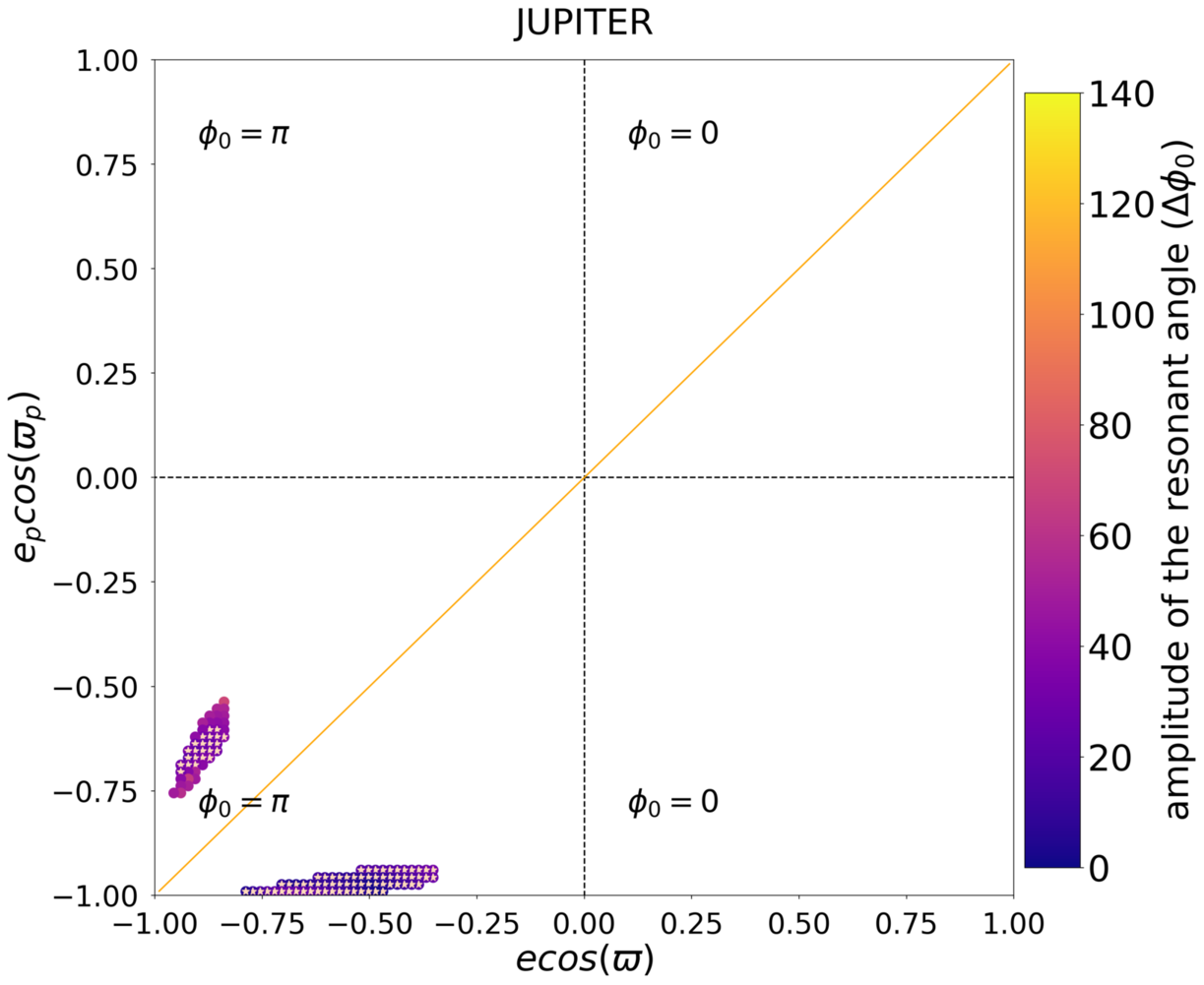}
         \caption{}
         \label{fig11:1jup}
     \end{subfigure}
     \vskip15pt
     \begin{subfigure}[b]{0.43\textwidth}
         \centering
         \includegraphics[width=\textwidth]{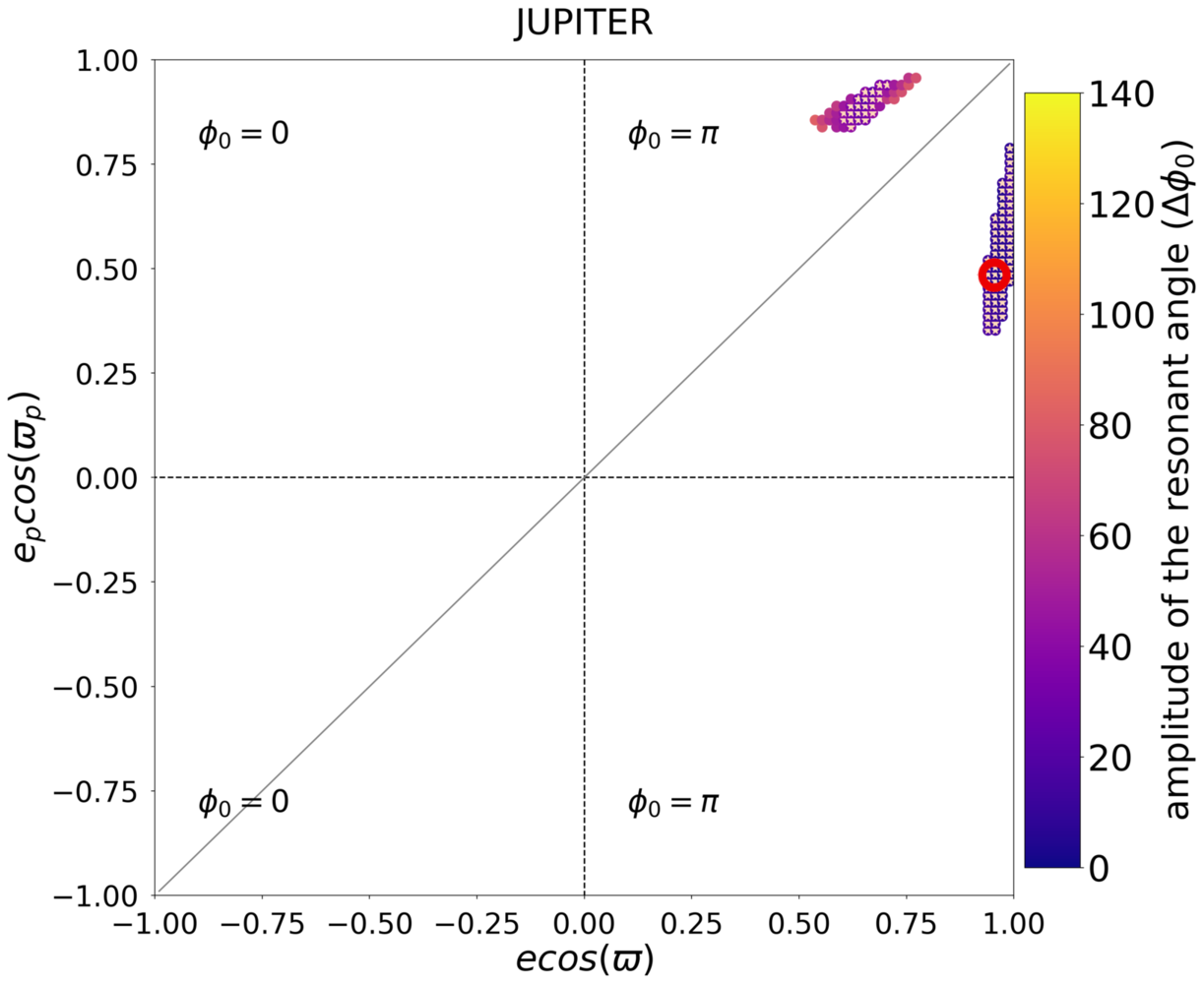}
         \caption{}
         \label{fig11:2jup}
     \end{subfigure}
     
     \caption{Resonant maps for the 1/-1 resonance in the planetary problem when the 2nd planet has Jupiter's mass (a) $M=0$; (b) $M=\pi$. The color bar represents the amplitude of the restricted angle ($\phi_{0}$) and the overlaying white symbols indicate the fixed point family where all resonant angles librate around a center. The orange and gray lines indicate collision at time zero or after half a period of the external object, respectively.}
     \label{fig:12}
\end{figure}

\begin{figure}
    \centering
     \includegraphics[width=0.43\textwidth]{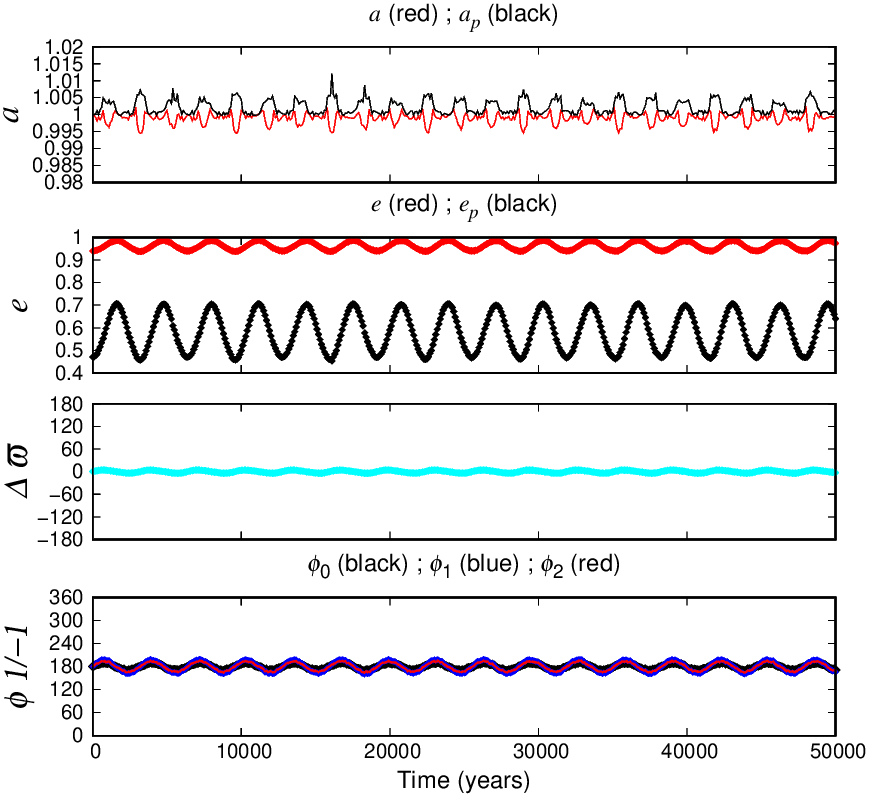}
    \caption{Orbital evolution as a function of time for the initial conditions circled in Figure  \ref{fig11:2jup} ($Q_1$, $M = \pi$).  The initial conditions are $e = 0.94$, $e_p = 0.47$. The 1st panel shows the semi-major axes of both planets, the 2nd panel shows their eccentricities, the 3rd panel shows the difference $\Delta\varpi$ between the longitudes of pericenter, the 4th panel shows the resonant angles $\phi_0$, $\phi_1$ and $\phi_2$.
    }
\label{fig:11jupci}
\end{figure}

The stability maps for the planetary problem when the 2nd planet has Jupiter's mass are presented in Figure \ref{fig:12}.  In this case the stable islands are associated with fixed point families which occur for the initial angle $\phi_0=\pi$ seen in $Q_3$ in Figure \ref{fig:12} (a) and in $Q_1$ in Figure \ref{fig:12} (b), while most  other  initial conditions lead to collision or escape.  Figure \ref{fig:11jupci} shows the orbital evolution of the initial condition marked by red circle in Figure \ref{fig:12} (b) ($e_0 = 0.94$, $e_p = 0.47$).  The  stability islands above and below the collision line at fixed $M=0$ or $M=\pi$ in \ref{fig:12} are not exactly symmetric (as we would expect as inverting the direction of motion of both planets results in the same relative motion) since they correspond to initial conditions with a time-lag of half a period (Figure \ref{fig:configurations}). This also explains the mirrored structures in quadrants $Q_3$ (Figure \ref{fig:12} (a)) and $Q_1$ (Figure \ref{fig:12} (a)). 

\begin{figure}
\centering
\begin{subfigure}[b]{0.43\textwidth}
    \includegraphics[width=\textwidth]{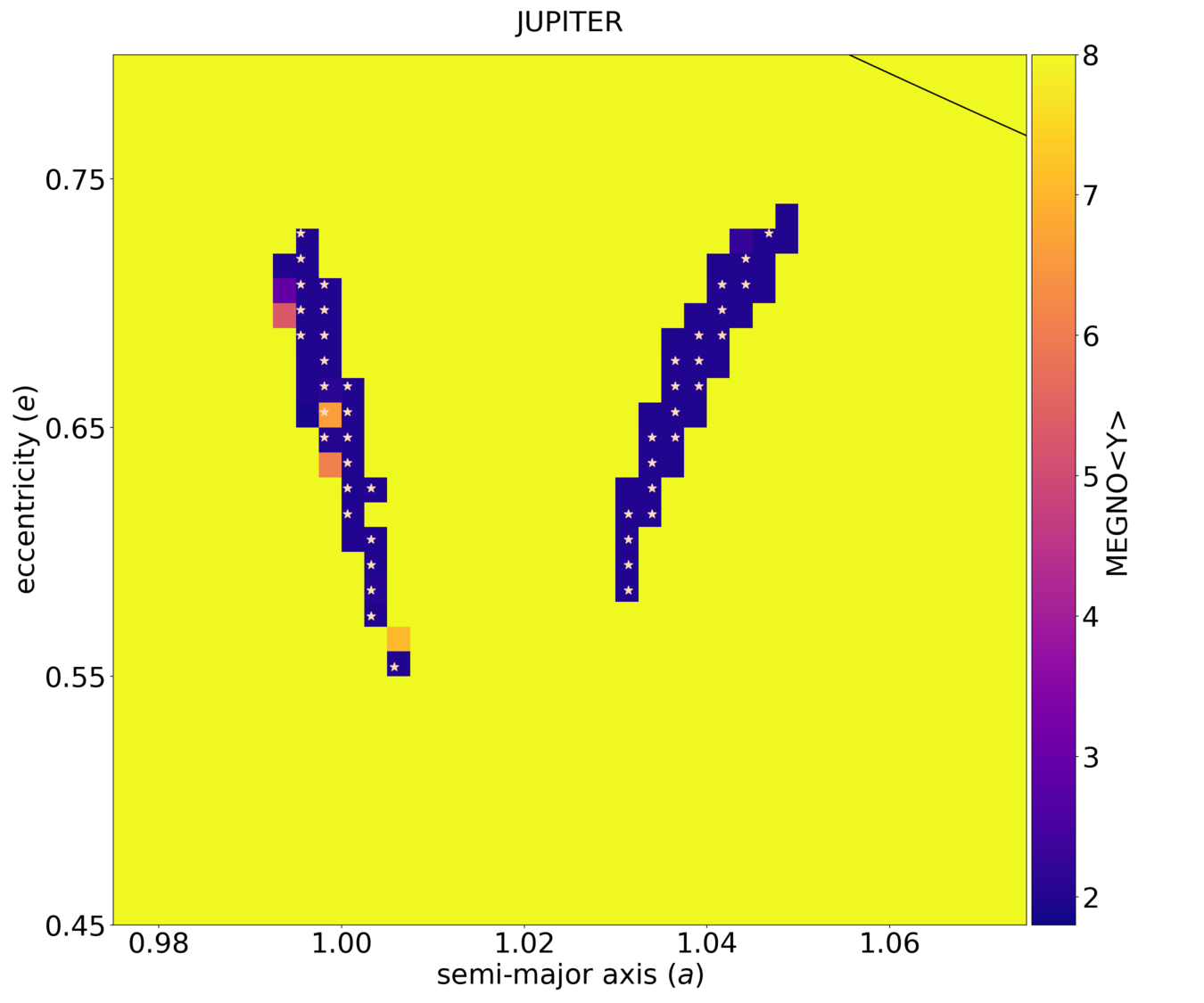}
    \caption{$e_p = 0.9$.}
    \label{fig:6meg}
\end{subfigure}
\vskip15pt
\begin{subfigure}[b]{0.43\textwidth}
     \includegraphics[width=\textwidth]{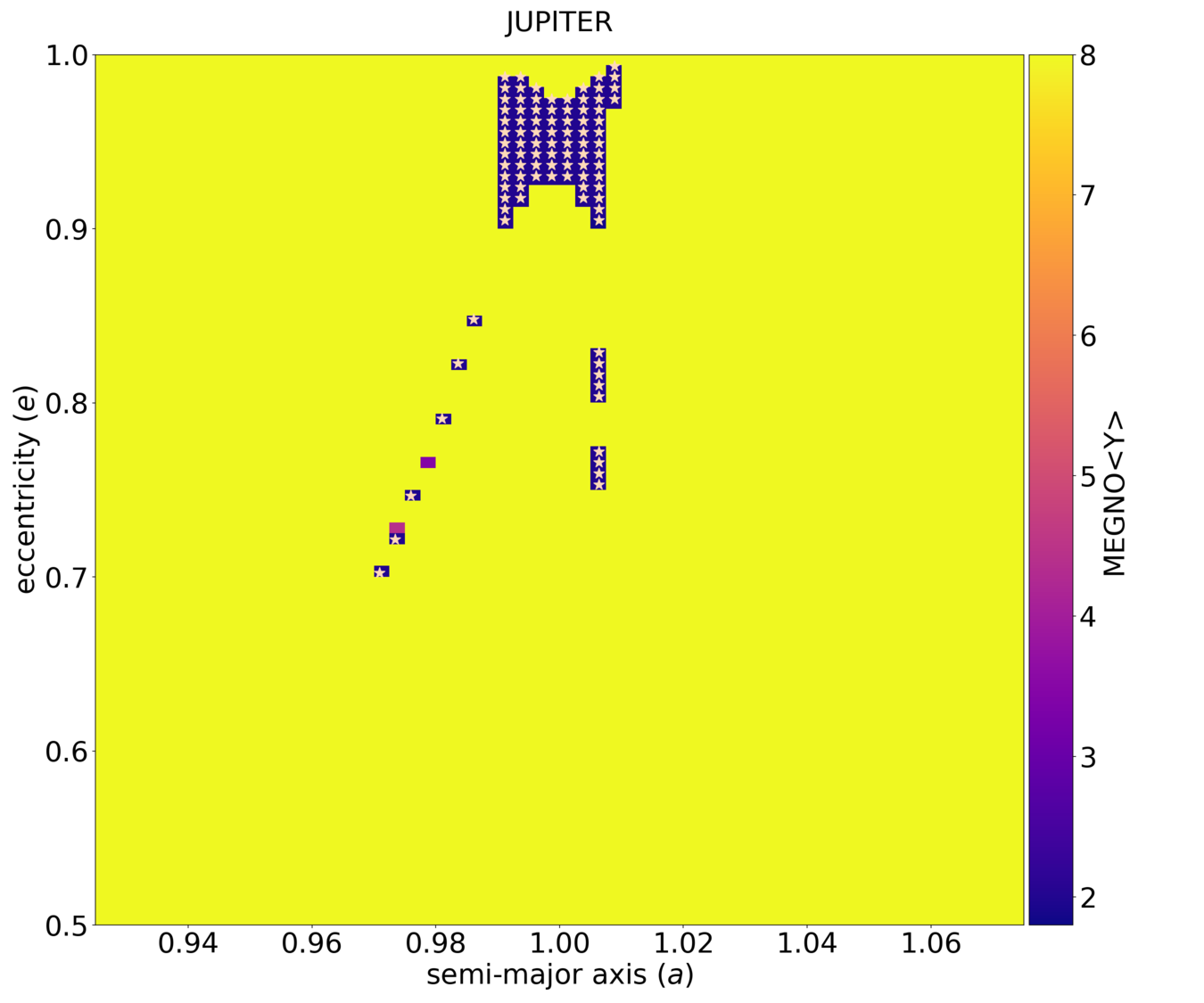}
    \caption{$e_p = 0.4$.}
    \label{fig:7meg}
\end{subfigure}

\caption{Stability map for resonance 1/-1 Jovian planetary system. The initial orbital elements adopted were $M = \pi$, $\omega=\omega_p=0$. (a) $e_p = 0.9$ (40x40 grid); (b) $e_p = 0.4$ (8x80 grid). The color bar represents the MEGNO value. The black line represents the collision line.}
\label{fig:megjup}
\end{figure}

We present $(a,e)$ stability maps in Figure \ref{fig:megjup} using initial conditions from  Figure \ref{fig:12}. In this case, the color bar  indicates the MEGNO value of the system, being stable blue/purple and unstable yellow, and the integration time span is $2\times 10^5\,T_p$. In \ref{fig:megjup} (a) we fix  $e_p = 0.9$  in order to explore the smaller fixed point island in Figure \ref{fig:12} (b). We see that in this case there are 2 branches of the fixed point family, one at $a = 1$, and the other at $a \approx 1.03$.  In \ref{fig:megjup} (a) we fix  $e_p = 0.4$  in order to explore the larger fixed point island in Figure \ref{fig:12} (b).  In this case the fixed point family is centered at $a=1$. 

\section{Discussion}

In the planar CR3BP  \citep{morais2013retrograde,morais2016retrograde,morais2016numerical,morais2019periodic,moraisetal2021} the periodic orbits associated with the resonant families  correspond  to fixed resonant angle  $\phi_{0}= -q\lambda - p\lambda_p + (p+q)\varpi$ \citep{morais2013retrograde}, being either 0 or $\pi$, with $\varpi$ circulating. We have seen here that in the ER3BP (non-zero eccentricity $e_p$) and in the planetary 3 body problem (2nd planet with non-zero mass) fixed point families of periodic orbits appear. Generally, as the mass ratio of the 2nd planet with respect to the 1st planet increases up to 1, the fixed point regions dominate the stable phase space. However, in the case of the 1/-2 and 2/-1 resonances, when the 2nd planet has Saturn's or Jupiter's mass, we saw that there are stable regions associated with libration of a mixed angle: $\phi_3 = -2\,\lambda - \lambda_p + \varpi_p + 2\varpi$ (1/-2); $\phi_3 = -\lambda - 2\lambda_p + 2\varpi_p + \varpi$ (2/-1). If we define the prograde planet to always be interior to the retrograde planet, then at the 2/-1 resonance $\phi_3 = -\lambda_p - 2\lambda + 2\varpi + \varpi_p$ which coincides with the definition of $\phi_3$ at the 1/-2 resonance. This is expected since if the exterior planet is in the 1/-2 resonance with the interior planet, then the interior planet is in the 2/-1 resonance with the exterior planet, and the resonant argument is the same.  As expected there is also exact symmetry between these configurations for the 1/-2 and 2/-1 resonances when the planets have identical masses.

\begin{figure}
        \centering
        \includegraphics[scale=0.14]{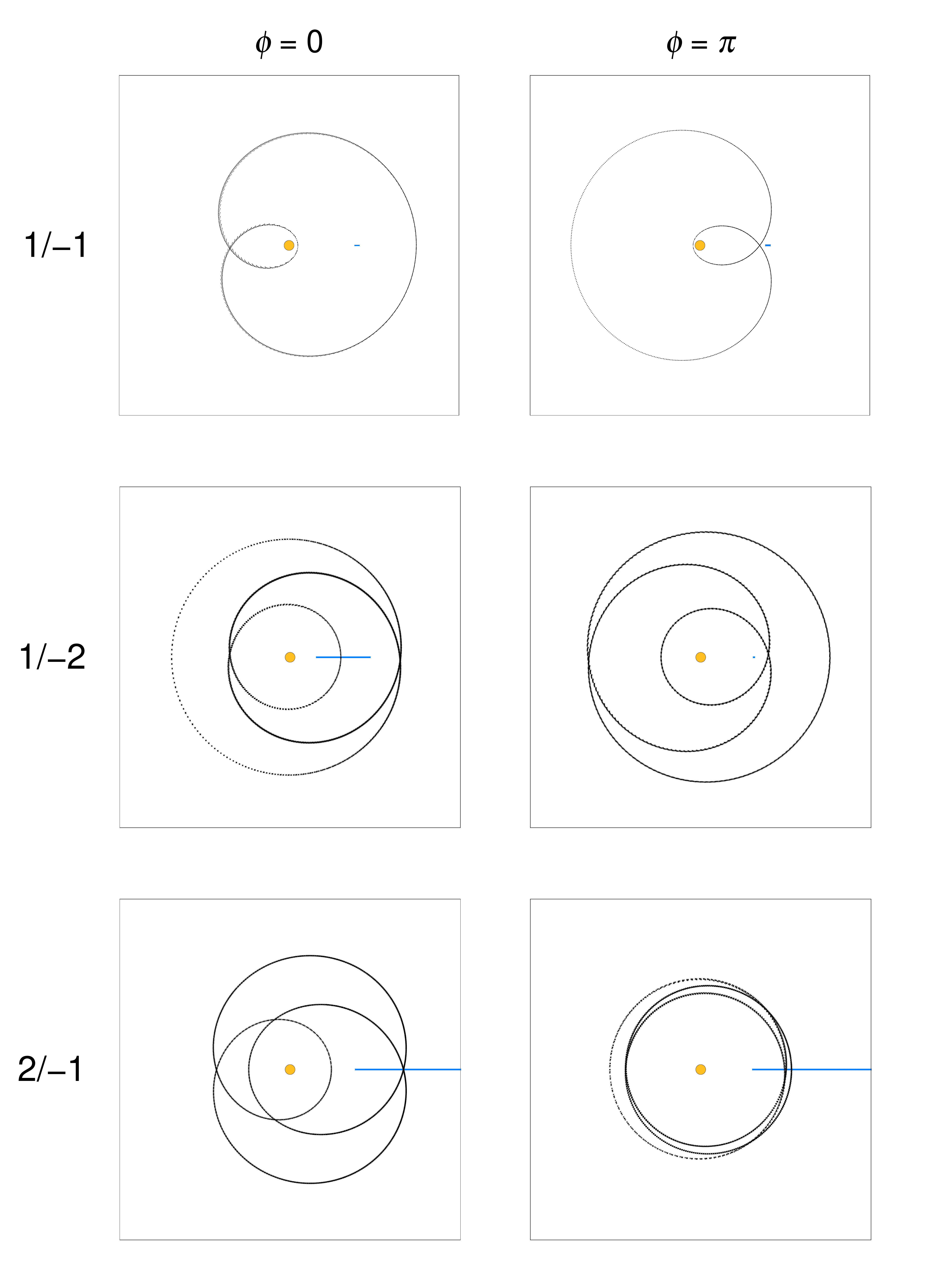}
        \vskip\baselineskip
        \caption{Orbits in synodic reference frame: 1/-1 (upper), 1/-2 (middle), 2/-1 (bottom) for $\phi = 0$ and $\pi$. The synodic orbit referred to the retrograde planet are represented in black. Except for the example of $2/-1$ for $\phi = \pi$ which is obtained using Jupiter's mass, the others are obtained using the mass of Neptune. All examples chosen are fixed points taken from the maps of the three previous chapters. The blue line represents the prograde planet orbit in a synodic reference frame which moves on the x-axis when $e_p \neq 0$.}    
        \label{fig:synodicorbits}
    \end{figure}

Figure \ref{fig:synodicorbits} shows periodic orbits of the planetary 3-body problem seen in the frame rotating with the prograde planet's true longitude angle,  as defined in \cite{hadjidemetriou20111}. They have similar shapes to the periodic orbits of the CR3BP for these resonances, except for the variation of the prograde's planet orbit as it moves between pericenter and apocenter, represented by the blue line on the x-axis
\citep{morais2013retrograde,morais2016retrograde}.

\section{Conclusion}

In this study we showed that there are stable configurations for the 1/2, 2/1 and 1/1 retrograde resonances in the planetary 3-body problem composed of a solar mass star, a Jupiter mass planet and a 2nd planet with either zero mass (ER3BP), or a non-zero mass equal to Neptune, Saturn or Jupiter. We saw that there are significant changes in the resonant phase space as we increase the mass ratio of the 2nd planet with respect to the 1st planet. 

In general, in the case of a 2nd planet with Neptune's mass, there are no significant differences in stability with respect to the ER3BP, except for the appearance of additional fixed point families within the stable regions.
As the mass of the 2nd planet increases the fixed point families become more predominant within the stable phase space. This happens for the 1/-2, 2/-1 and 1/-1 resonances.

The  1/1 retrograde resonance in the E3RBP has stable regions of quasi-periodic orbits associated with the CR3BP resonant centers $\phi_0=0$ and $\phi_0=\pi$, and an additional family of fixed point periodic orbits appears at large eccentricity of the prograde planet, $e_p$.     
There are also stable retrograde coorbital configurations in the planetary 3-body problem. As the 2nd planet's mass increases, the $\phi_0=0$ center becomes less stable than the $\phi_0=\pi$ center, which occurs at large $e$. In the case of 2 Jovian mass planets the stable regions correspond to fixed point periodic families with $\phi_0=\pi$ and $\varpi-\varpi_p=0$,  that occur at large eccentricities. 

The results presented in this article depend on the mass ratios and relative distances only hence may be applied to other systems. It has been proposed that counter revolving resonant planetary systems may exist around other stars \citep{gayon2008retrograde,gayon2009fitting}.  Our results show which stable configurations are possible and therefore may guide searches for such systems.

\section*{Acknowledgements}

This work was funded  by the Coordenação de Aperfeiçoamento de Pessoal de Nível Superior – Brasil (CAPES) – Finance Code 001. The authors acknowledge support from Grants FAPESP/2019/24958-5 \& FAPESP/2021/11982-5 of São Paulo Research Foundation and from CNPQ-Brazil (PQ2/304037/2018-4). This work used computational resources supplied by the Center for Scientific Computing (NCC/GridUNESP) of the São Paulo State University (UNESP). 

\section*{Data Availability}

The data underlying this paper will be shared on reasonable request
to the corresponding author.



\bibliographystyle{mnras}
\bibliography{example} 

\begin{thebibliography}{}
\makeatletter
\relax
\def\mn@urlcharsother{\let\do\@makeother \do\$\do\&\do\#\do\^\do\_\do\%\do\~}
\def\mn@doi{\begingroup\mn@urlcharsother \@ifnextchar [ {\mn@doi@}
  {\mn@doi@[]}}
\def\mn@doi@[#1]#2{\def\@tempa{#1}\ifx\@tempa\@empty \href
  {http://dx.doi.org/#2} {doi:#2}\else \href {http://dx.doi.org/#2} {#1}\fi
  \endgroup}
\def\mn@eprint#1#2{\mn@eprint@#1:#2::\@nil}
\def\mn@eprint@arXiv#1{\href {http://arxiv.org/abs/#1} {{\tt arXiv:#1}}}
\def\mn@eprint@dblp#1{\href {http://dblp.uni-trier.de/rec/bibtex/#1.xml}
  {dblp:#1}}
\def\mn@eprint@#1:#2:#3:#4\@nil{\def\@tempa {#1}\def\@tempb {#2}\def\@tempc
  {#3}\ifx \@tempc \@empty \let \@tempc \@tempb \let \@tempb \@tempa \fi \ifx
  \@tempb \@empty \def\@tempb {arXiv}\fi \@ifundefined
  {mn@eprint@\@tempb}{\@tempb:\@tempc}{\expandafter \expandafter \csname
  mn@eprint@\@tempb\endcsname \expandafter{\@tempc}}}

\bibitem[\protect\citeauthoryear{Chambers}{Chambers}{1999}]{chambers1999hybrid}
Chambers J.~E.,  1999, Monthly Notices of the Royal Astronomical Society, 304,
  793

\bibitem[\protect\citeauthoryear{Cincotta \& Sim{\'o}}{Cincotta \&
  Sim{\'o}}{2000}]{cincotta2000simple}
Cincotta P.~M.,  Sim{\'o} C.,  2000, Astronomy and Astrophysics Supplement
  Series, 147, 205

\bibitem[\protect\citeauthoryear{{Dobrovolskis}}{{Dobrovolskis}}{2012}]{2012DPS....4411222D}
{Dobrovolskis} A.~R.,  2012, in AAS/Division for Planetary Sciences Meeting
  Abstracts \#44. p. 112.22

\bibitem[\protect\citeauthoryear{Gayon \& Bois}{Gayon \&
  Bois}{2008}]{gayon2008retrograde}
Gayon J.,  Bois E.,  2008, Astronomy \& Astrophysics, 482, 665

\bibitem[\protect\citeauthoryear{Gayon-Markt \& Bois}{Gayon-Markt \&
  Bois}{2009}]{gayon2009fitting}
Gayon-Markt J.,  Bois E.,  2009, Monthly Notices of the Royal Astronomical
  Society: Letters, 399, L137

\bibitem[\protect\citeauthoryear{Go{\'z}dziewski}{Go{\'z}dziewski}{2003}]{gozdziewski2003stability}
Go{\'z}dziewski K.,  2003, Astronomy \& Astrophysics, 398, 1151

\bibitem[\protect\citeauthoryear{Hadjidemetriou \& Voyatzis}{Hadjidemetriou \&
  Voyatzis}{2011}]{hadjidemetriou20111}
Hadjidemetriou J.~D.,  Voyatzis G.,  2011, Celestial Mechanics and Dynamical
  Astronomy, 111, 179

\bibitem[\protect\citeauthoryear{Kotoulas \& Voyatzis}{Kotoulas \&
  Voyatzis}{2020a}]{kotoulas2020retrograde}
Kotoulas T.,  Voyatzis G.,  2020a, Celestial Mechanics and Dynamical Astronomy,
  132, 1

\bibitem[\protect\citeauthoryear{Kotoulas \& Voyatzis}{Kotoulas \&
  Voyatzis}{2020b}]{kotoulas2020planar}
Kotoulas T.,  Voyatzis G.,  2020b, Planetary and Space Science, 182, 104846

\bibitem[\protect\citeauthoryear{Kotoulas, Voyatzis  \& Morais}{Kotoulas
  et~al.}{2022}]{kotoulas2022three}
Kotoulas T.,  Voyatzis G.,   Morais M. H.~M.,  2022, Planetary and Space
  Science, 210, 105374

\bibitem[\protect\citeauthoryear{Malmberg, Davies  \& Heggie}{Malmberg
  et~al.}{2011}]{malmberg2011effects}
Malmberg D.,  Davies M.~B.,   Heggie D.~C.,  2011, Monthly Notices of the Royal
  Astronomical Society, 411, 859

\bibitem[\protect\citeauthoryear{Morais \& Giuppone}{Morais \&
  Giuppone}{2012}]{morais2012stability}
Morais M.,  Giuppone C.,  2012, Monthly Notices of the Royal Astronomical
  Society, 424, 52

\bibitem[\protect\citeauthoryear{Morais \& Namouni}{Morais \&
  Namouni}{2013a}]{morais2013retrograde}
Morais M.,  Namouni F.,  2013a, Celestial Mechanics and Dynamical Astronomy,
  117, 405

\bibitem[\protect\citeauthoryear{Morais \& Namouni}{Morais \&
  Namouni}{2013b}]{morais2013asteroids}
Morais M.,  Namouni F.,  2013b, Monthly Notices of the Royal Astronomical
  Society: Letters, 436, L30

\bibitem[\protect\citeauthoryear{Morais \& Namouni}{Morais \&
  Namouni}{2016a}]{morais2016retrograde}
Morais M.,  Namouni F.,  2016a, Computational and Applied Mathematics, 35, 881

\bibitem[\protect\citeauthoryear{Morais \& Namouni}{Morais \&
  Namouni}{2016b}]{morais2016numerical}
Morais M. H.~M.,  Namouni F.,  2016b, Celestial Mechanics and Dynamical
  Astronomy, 125, 91

\bibitem[\protect\citeauthoryear{Morais \& Namouni}{Morais \&
  Namouni}{2017}]{morais2017reckless}
Morais H.,  Namouni F.,  2017, Nature, 543, 635

\bibitem[\protect\citeauthoryear{Morais \& Namouni}{Morais \&
  Namouni}{2019}]{morais2019periodic}
Morais M.,  Namouni F.,  2019, Monthly Notices of the Royal Astronomical
  Society, 490, 3799

\bibitem[\protect\citeauthoryear{Morais, Namouni, Voyatzis  \& Kotoulas}{Morais
  et~al.}{2021}]{moraisetal2021}
Morais M.,  Namouni F.,  Voyatzis G.,   Kotoulas T.,  2021, Celestial Mechanics
  and Dynamical Astronomy, 133, 1

\bibitem[\protect\citeauthoryear{Namouni \& Morais}{Namouni \&
  Morais}{2015}]{namouni2015resonance}
Namouni F.,  Morais M. H.~M.,  2015, Monthly Notices of the Royal Astronomical
  Society, 446, 1998

\bibitem[\protect\citeauthoryear{Namouni \& Morais}{Namouni \&
  Morais}{2018}]{namouni2018coorbital}
Namouni F.,  Morais H.,  2018, Computational and Applied Mathematics, 37, 65

\bibitem[\protect\citeauthoryear{Rein \& Spiegel}{Rein \&
  Spiegel}{2015}]{rein2015ias15}
Rein H.,  Spiegel D.~S.,  2015, Monthly Notices of the Royal Astronomical
  Society, 446, 1424

\bibitem[\protect\citeauthoryear{Wiegert, Connors  \& Veillet}{Wiegert
  et~al.}{2017}]{wiegert2017retrograde}
Wiegert P.,  Connors M.,   Veillet C.,  2017, Nature, 543, 687

\makeatother
\end{thebibliography}



\appendix


\clearpage
\bsp	
\label{lastpage}
\end{document}